\documentclass{article}
\pdfoutput=1

\usepackage[utf8]{inputenc}
\usepackage{jinstpub}
\usepackage{amsmath}
\usepackage{bm}
\usepackage[titlenumbered, ruled]{algorithm2e}

\DeclareMathOperator*{\argmin}{arg\,min}

\title{Nonnegative Gaussian process tomography for generalized segmented planar detectors}
\author[a,c,1]{D.~Blyth\note{Corresponding Author}}
\author[a,b]{N.~Mullins}
\author[a]{E.~Galyaev}
\author[a,b]{J.~Holmes}

\affiliation[a]{Radiation Detection and Imaging (RDI), LLC\\
1801 S Jentilly Ln Ste B-18, Tempe, AZ 85281-5759, USA}
\affiliation[b]{Department of Physics, Arizona State University\\
550 E Tyler Drive, Tempe, AZ, USA}
\affiliation[c]{Phynix Technologies, LLC\\
8508 E Appaloosa Trl, Scottsdale, AZ 85258-1483, USA}

\emailAdd{bej1eu@gmail.com}

\abstract{The concept of Gaussian process tomography along with nonnegative constraints is applied in the context of high-resolution image reconstruction using segmented planar detectors with few readout channels.  Expanding on the concept of 2-D projections onto strip-like readout segmentations, 3-D projections as well as more generalized detector segmentation and readout channel mappings are explored.  A focus is placed on reconstructing dose distributions in proton therapy pencil beam scanning, and a fast, approximate approach to applying nonnegative constraints is developed and motivated for use in proton therapy beam imaging.}


\begin{document}

\maketitle

\section{Introduction}
\begin{figure}
    \centering
    \includegraphics[width=4.5in]{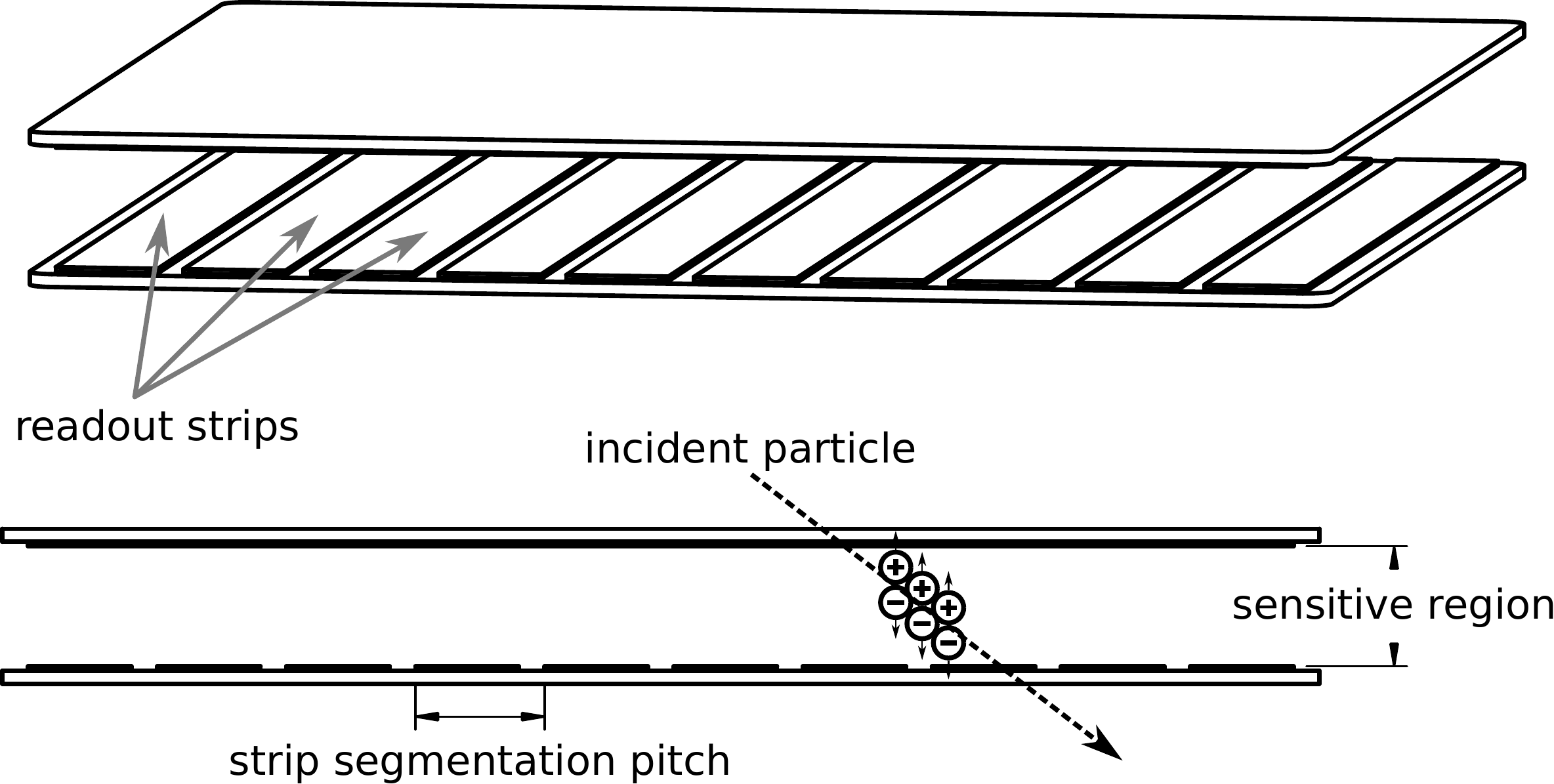}
    \caption{\textbf{Strip segmentation of planar detector readout} - Shown is an illustration of strip segmentation of the anode of a particle detector from an angle (top) and from the side (bottom).  A particle transition deposits energy and induces separation of positive and negative charge carriers, and an applied electric field causes the charges to drift towards the anode and cathode.  We define the term segmentation herein as a particular pattern of isolated regions of the anode and/or cathode and their connection to electronic readout channels.}
    \label{figSegmentation}
\end{figure}

Detectors for measuring the planar distributions of proton therapy fluence are often pixelated in their readout, and must also be relatively large-area to accommodate a wide range of beam positions (e.g., \cite{det_matrixx_1, stelljes2015dosimetric}).  Since the required number of readout channels for pixelated detectors with a fixed resolution requirement goes linearly with area, a large detector typically suffers from having either slow (multiplexed) readout or poor resolution.  It is a common approach in physics research to solve this problem --~which one may identify as a separation between detector scale and the required resolution scale~-- by making independent linear projections of the detector signal onto strip-like planar readout geometries (see Figure \ref{figSegmentation}).  This approach takes advantage of the sparse distributions of charge in the detector, improving resolution for a given number of readout channels while introducing acceptable degeneracy in the space of possible signal distributions.

Typically, one may resolve introduced degeneracies by applying physical knowledge of the measurement.  In medium- and high-energy physics, the knowledge that is applied is usually that the signals of interest come from single particle tracks.  Alternatively we can consider a lower bandwidth or higher intensity scenario, where we observe a distribution of particle tracks at any point in time, and apply physical knowledge about that distribution.  Taking scanning-beam proton therapy for example, it's reasonable to assume the proton flux to be distributed as a covariant 2-D Gaussian when integrated over time scales on the order of milliseconds or less (\cite{shen2017using, kohno2017development, pidikiti2018commissioning} give examples of spot scan timing).  Within the 2-D Gaussian assumption, two orthogonal readout projections (strip-like readout) could be used to measure the proton flux, but this would allow degeneracy in the covariance of the distribution.  The use of three linearly-independent projections, however, would provide information on the full mean position and covariance matrix \cite{fraser1979beam}, thus providing an acceptable level of degeneracy that is resolved by the assumption of a covariant 2-D Gaussian distribution.  Therefore, a fast detector (with a bandwidth of $>1$~kHz) with three strip-like independent projections can reconstruct the overall dose distribution for scanning-beam proton therapy with high precision and no loss of generality (within the 2-D Gaussian assumption).

In this work, we explore an approach that generalizes the reconstruction of an image representing particle flux through a detector from arbitrary planar electrode segmentation, where the reconstruction is performed on a short enough time scale to be useful for live imaging of radiotherapy beams.  This image could be used directly or, e.g., as an intermediate step for calculating moments.  We adopt the technique of Gaussian process tomography (GPT) \cite{svensson2011non}, inspired by use cases in fusion research for reconstructing plasma distributions from soft x-rays \cite{li2013bayesian, wang2018gaussian}.  GPT allows the image reconstruction to be formulated within the framework of generalized $\chi^2$ minimization, where pixels in the image are treated as parameters.  Such an optimization is typically ill-conditioned by itself, since the number of virtual pixels is typically much larger than the number of readout channels.  However, the GPT concept introduces a Bayesian prior on the pixel covariance matrix using a squared exponential kernel applied to the image coordinate space.  This prior assumption about the probability distribution of pixel values acts as a regularization term in the objective function for fit optimization that resolves the indeterminacy of the fit with regularization parameters that can be physically motivated \emph{a priori}.  Additionally, we apply important inequality constraints to the virtual pixel optimization in order to enforce the physical constraint that fluence must be nonnegative.

\section{Tomography formulation}
\label{secGPT}
Consider a planar radiation detector; an example of which is illustrated in Figure \ref{figSegmentation}.  This could be a gas or semiconductor with electrodes on either side of a broad area and an applied bias.  When energy is deposited in the gas or depletion region, charges are separated in pairs of electrons/ions or electrons/holes, which drift towards opposing electrodes.  While drifting, the charge carriers also diffuse.  Our aim is to reconstruct the distribution of power deposited into the detector gas/depletion region, projected onto a plane parallel to the electrode planes.

We consider this distribution of power deposited by incident radiation to come from a particular Gaussian process (GP).  A GP is defined as a collection of random variables, where any finite subset forms a joint Gaussian distribution.  The GP is effectively defined by a covariance matrix generating function (or kernel), where in this case we choose what is referred to as the squared exponential (SE) function
\begin{equation}
\label{eqSEKernel}
    k_\text{SE}\left(x, x'\right) = \exp{\frac{-\left|x-x'\right|^2}{2l^2}},
\end{equation}
where $x$ and $x'$ are coordinates of the distribution. From this GP, we take a rectangular grid of points that we call virtual pixels, and use the kernel to generate a covariance matrix.  Additionally, in order to keep calculations involving this covariance matrix tractable by describing it as a sparse matrix, we truncate the kernel as seen below.
\begin{equation}
\label{eqSEKernelCutoff}
    k^{\text{cutoff}}_\text{SE}\left(x, x'\right) =
    \begin{cases}
        k_\text{SE}\left(x, x'\right) & k_\text{SE}\left(x, x'\right) > \text{cutoff threshold} \\
        0 & k_\text{SE}\left(x, x'\right) \leq \text{cutoff threshold}
    \end{cases}
\end{equation}
Along with an assumption of the mean values and overall covariance scaling factor, this kernel choice forms a Bayesian prior probability distribution for the virtual pixels which we can update using detector data.  We choose a mean value assumption of zero.

In order to understand how to update the virtual pixel prior using detector data, we form a linear equation relating depositions at each virtual pixel to current extracted at sensitive electrodes.\footnote{For the case of proton beam therapy, changes in the incident particle flux occur on time scales that are long compared to the drift time of the charged particle carriers in a detector with a reasonable drift gap.  These conditions allow us to ignore transient induced currents on neighboring electrodes.}
We define this linear relationship as
\begin{equation}
\label{eqLinearRelationship}
    d = G\Phi + \Delta,
\end{equation}
where $d$ is a vector of length $m$ containing the current measurements of all sensitive electrodes, $\Phi$ is a vector of length $n$ containing values proportional to the true power deposited at each virtual pixel, $G$ is an $m\times n$ design matrix that describes the relationship in the absence of noise, and $\Delta$ is a vector representing stochastic noise in the measurement.  For our purposes, $G$ contains information about the electrode segmentation and charge carrier diffusion.\footnote{The inclusion of diffusion in $G$ implies the assumption that incident radiation is perpendicular to the electrode and reconstruction planes.  If this is not a valid assumption, diffusion cannot be included here, or reconstruction must be extended to include depth within the detector's sensitive volume.}

\begin{figure}
    \centering
    \includegraphics[width=1.5in]{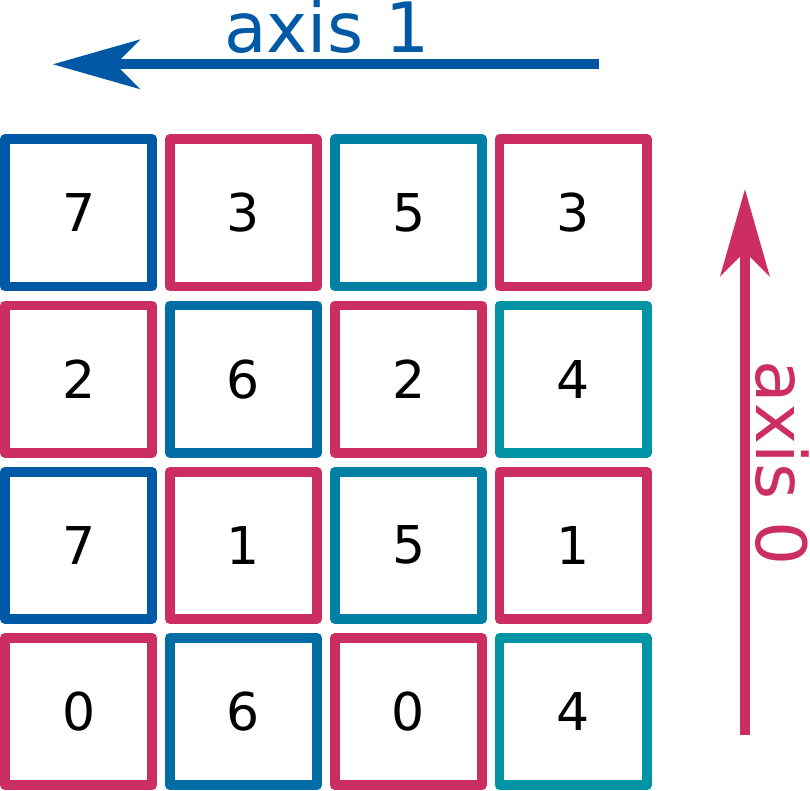}
    \hspace{0.5in}
    \includegraphics[width=1.5in]{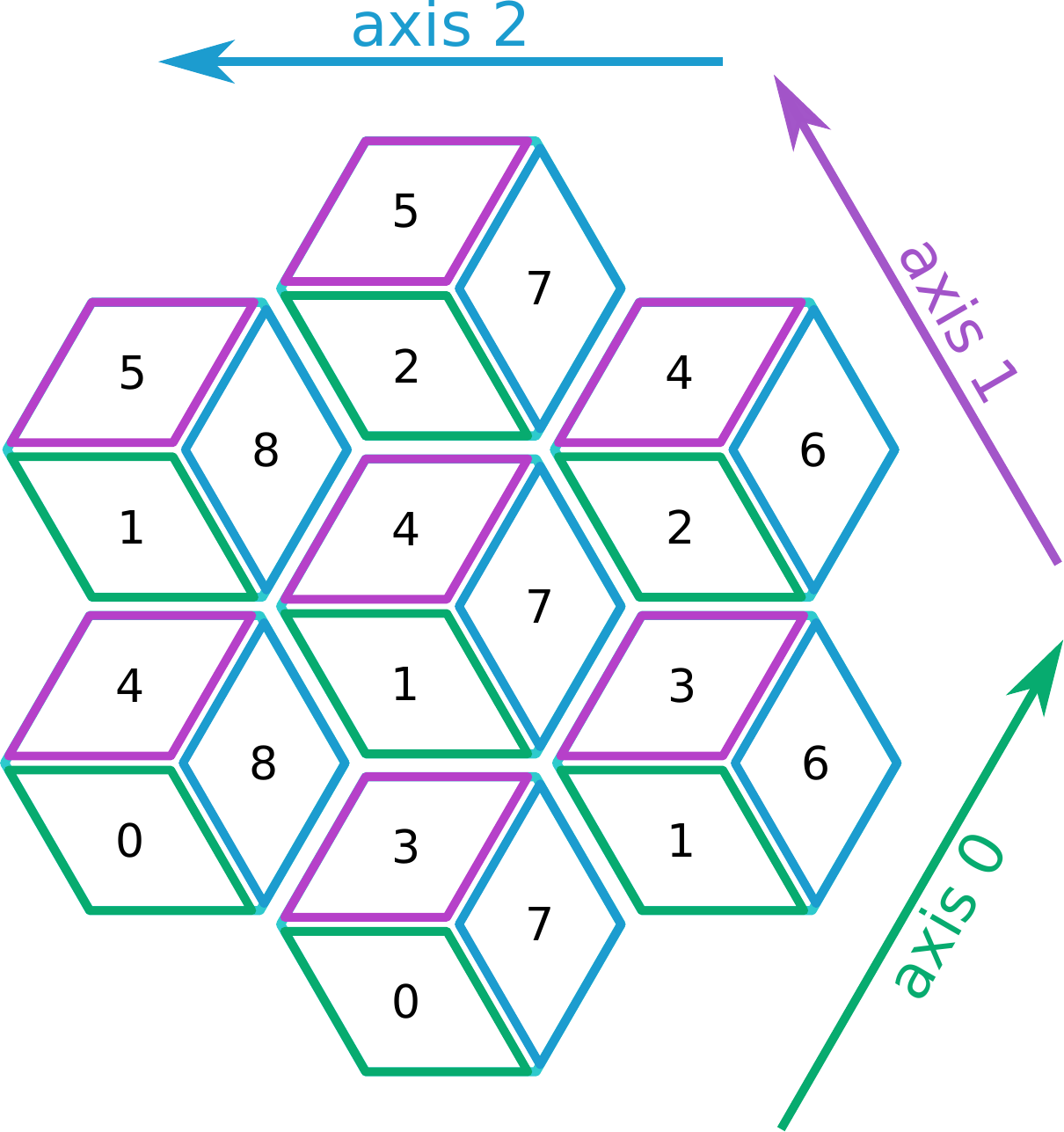}
    \caption{\textbf{Example 2-D and 3-D electrode segmentation} - Shown are two examples of electrode segmentation schemes for 2-D (left) and 3-D (see \cite{rdi_patent_one}) (right) projections in a single plane.  The color indicates a particular projection, while the number in the center of each pad indicates an assigned channel number.  The assignment of the same channel to multiple pads implies a connection between them established by, e.g., printed circuit board vias and traces.  In both cases, the pad layout is regular such that the pad that is effectively ``below'' a point in reconstruction space can be found by finding the nearest pad center.}
    \label{figExampleReadout}
\end{figure}

To generate $G$, a Monte Carlo approach is taken, where random coordinates $x$ and $y$ (samples) are taken surrounding each virtual pixel center, combined with random sampling of diffusion (further detail on diffusion in section \ref{subsecDiffusion}).  For each sample, we use a $k$-D tree lookup to determine the nearest pad and corresponding readout channel in a regular array of pads that make up a segmented electrode plane (see Figure~\ref{figExampleReadout} for examples of electrode segmentation), and bin normalized samples into the corresponding element of $G$:
\begin{equation}
\label{eqDesignMatrix}
    G_{ij} = \frac{1}{N}\sum^{N}{f_i\left(x,y\sim\mathcal{U}\left(\text{bounds for pixel }j\right)+\text{diffusion}\right)},
\end{equation}
where  $f$ yields a one-hot vector with 1 at the index of the nearest channel, and 0 elsewhere.  Using this approach, we can fill $G$ with samples at a rate of $\sim 10^6$ per second, and typically complete the process in 1 to 200~s, depending on the required image size and sample density.  $G$ only needs to be calculated once for each detector geometry.

For subsections \ref{subsecGLS}, \ref{subsecGPTReg}, and \ref{subsecNonneg}, it should be noted that the above assertions about planar geometry and the inclusion of diffusion need not apply.  The formulation presented is applicable anywhere that equation~(\ref{eqLinearRelationship}), the GP prior, and nonnegativity apply.

\subsection{Generalized least squares}
\label{subsecGLS}
Since in reality $\Phi$ and $\Delta$ cannot be known exactly, we reformulate the relationship in terms of the estimated virtual pixel values $\phi$ and the residuals $\delta$,
\begin{equation}
    \delta = d - G\phi,
\end{equation}
and write down the $\chi^2$ value as the the Mahalanobis magnitude of the residuals (proportional to the log of the likelihood function),
\begin{equation}
\label{eqChiSquareLS}
    \chi_\text{LS}^2 \equiv \left(d-G\phi\right)^\top\Sigma_d^{-1}\left(d-G\phi\right),
\end{equation}
where $\Sigma_d$ is the covariance matrix for the measurements.  Minimizing $\chi_\text{LS}^2$ from the above equation~(w.r.t. $\phi$) is equivalent to maximizing the likelihood function for $\phi$.

In the absence of constraints, equation~(\ref{eqChiSquareLS}) can be minimized in the typical way, i.e. by setting each component of the gradient $\nabla_\phi \chi_\text{LS}^2$ equal to zero.  The solution for $\phi$ is straightforward to find:
\begin{equation}
    \phi = \left(G^\top \Sigma_d^{-1}G\right)^{-1}G^\top \Sigma_d^{-1}d.
\end{equation}
Upon consideration, two significant practical problems with this solution can be found.  Firstly, the matrix $G^\top \Sigma_d^{-1}G$ is in general not full rank, and is guaranteed to not be full rank if the number of readout channels $m$ is less than the number of virtual pixels $n$.  Consequently, there is no unique solution to $\phi$ for cases we are interested in.  Secondly, inversion of the matrix in question is performed on the (potentially very large) virtual pixel space, and can quickly become intractable.  For context, in section \ref{secSegmentationStudies} we will consider cases where $n=10^4$.  Recall that the computation time of a naive matrix inversion is $\mathcal{O}\left(n^3\right)$.

\subsection{GPT regularization}
\label{subsecGPTReg}
In order to physically motivate a unique solution for $\phi$, we adopt the concept of GPT \cite{svensson2011non}.  At this point, GPT can be described quite simply as the application of a Gaussian process Bayesian prior assumption for $\phi$.  Applying Bayes' theorem to update our prior with detector data, we have
\begin{equation}
    p\left(\phi|d\right) \propto p\left(d|\phi\right) p\left(\phi\right),
\end{equation}
where $p\left(d|\phi\right)$ is our likelihood function and $p\left(\phi\right)$ is our prior assumption of $\phi$.  With zero mean on the prior distribution, our $\chi^2$ objective function gains an added quadratic regularization term proportional to the Mahalanobis magnitude of $\phi$:
\begin{equation}
\label{eqChiSquareGPT}
    \chi_\text{GPT}^2 = \chi_\text{LS}^2 + \sigma_p^{-1}\phi^\top \Sigma_p^{-1}\phi,
\end{equation}
where $\Sigma_p$ is the covariance matrix generated from the virtual pixel coordinates and the generating function in equation~(\ref{eqSEKernelCutoff}), and $\sigma_p$ is a scalar weight that subsumes the scale of the covariance matrix.

Setting the gradient $\nabla_\phi\chi_\text{GPT}^2$ to zero, we end up with updated mean values
\begin{equation}
\label{eqPhiGPT}
    \phi = \left(\sigma_p^{-1}\Sigma_p^{-1} + G^\top \Sigma_d^{-1}G\right)^{-1}G^\top \Sigma_d^{-1}d.
\end{equation}
We can see that the GPT regularization term has turned our problem from being ill-conditioned to performing a rank update to an inverse that is already known ($\Sigma_p$).  However, we are still left with the problem of computational tractability.  Fortunately, equation~(\ref{eqPhiGPT}) is a clear application for the Woodbury matrix identity, which we can rewrite as
\begin{equation}
\label{eqWoodburyIdentity}
    \left(\sigma_p^{-1}\Sigma_p^{-1} + G^\top \Sigma_d^{-1}G\right)^{-1} = \sigma_p\Sigma_p - \sigma_p^2\Sigma_pG^\top\left(\Sigma_d + G\Sigma_pG^\top\right)^{-1}G\Sigma_p,
\end{equation}
where the matrix inversion has been moved from the large virtual pixel space to the typically small readout channel space.  Defining the constant matrix
\begin{equation}
\label{eqADef}
    A \equiv \Sigma_pG^\top
\end{equation}
and simplifying, we end up with
\begin{equation}
\label{eqGPTSolution}
\begin{split}
    \phi & = A\left(\Sigma_d + \sigma_pGA\right)^{-1}d\\
    & = A\left[\left(\Sigma_d + \sigma_pGA\right)\setminus d\right].
\end{split}
\end{equation}
In the second line of the above equation, we use the convention that $M \setminus b$ is the solution to $x$ for $Mx = b$.  A direct solution as opposed to performing an inversion is numerically stable, and (especially since $\Sigma_d + \sigma_pGA$ can be decomposed by Cholesky factorization) it is typically faster to decompose and direct solve than it is to invert.

Note that there are two parameters of the GPT regularization that have thus far been left unmotivated: $l$ from equation~(\ref{eqSEKernel}) and $\sigma_p$ from equation~(\ref{eqChiSquareGPT}).  Ideally, these would be carefully configured to provide the optimal image reconstruction (previous works have applied the Bayesian Occam's razor formalism).  However, we find that these parameters can be easily motivated by knowledge of the measurement.  The length scale $l$ can be interpreted roughly as a regularization cutoff scale, and should be set near the scale of the smallest features of interest.  The regularization weighting factor $\sigma_p$ must be set large enough to minimize the bias towards zero introduced by the Bayesian prior.  In practice, we find that $\sigma_p$ can be set effectively based on the scale of the detector signal immediately prior to reconstruction.  Since $\sigma_p$ is the width of the prior univariate distribution for a single virtual pixel, $\sigma_p$ should typically be greater than the max anticipated pixel value.

\subsection{Nonnegativity}
\label{subsecNonneg}
One would hope that equation~(\ref{eqGPTSolution}) would be sufficient for useful image reconstruction, due to the fact that exact solutions can be found very quickly (we have found that it can be done in in less than a millisecond, depending on the configuration).  Additionally, an approximate approach where $\Sigma_d$ is assumed for all time can be performed by a single $n\times m$ matrix multiplication on $d$, or can be reduced further ahead of time to moments in pixel coordinate space (such as the mean and covariance of the image) that take on the order of a microsecond to calculate.  While this sort of fast mean moment has been found to give reasonable results, they are relatively noisy, and higher moments like covariance tend to give poor results.  These poor results are largely due to the fact that pixel values are allowed to take on nonphysical negative values.

While there is definitely room here for creative solutions like using additional detector information to make better prior assumptions of the mean pixel values, we have taken a brute force approach for the sake of generality, which is to apply nonnegativity constraints to the pixels.  Additionally, for the sake of speed we chose to base our approach on so-called active set approaches to nonnegative least squares \cite{lawson1995solving, chen2010nonnegativity}, rather than taking a Gibbs sampling approach as is done in \cite{svensson2011non}.  Active set approaches apply nonnegativity constraints by attempting to identify the set of actively constrained pixels in a stepwise fashion and treat them as equality constraints.

\paragraph{KKT conditions}  The Karush, Kuhn, Tucker (KKT) conditions are general first-derivative necessary conditions for optimality under given sets of equality and inequality constraints.  These conditions can be used as tests to guide an active set search.  There are four types of conditions called stationarity (combined constraints and objective gradient must be parallel), primal feasibility (equality and inequality constraints must be satisfied), dual feasibility (inequality constraints can only penalize the objective function), and complementary slackness (inequality constrained parameters need not have a zero gradient).  Applying the KKT conditions yields the following three sets of tests.
\begin{equation}
\label{eqKKTConditions}
\begin{split}
    \phi_i & \geq 0 \text{ for } i\in\{0,...,n-1\} \text{ (primal feasibility)}\\
    \left(\nabla_\phi\chi_\text{GPT}^2\right)_i & \geq 0 \text{ for } i\in\{0,...,n-1\} \text{ (dual feasibility)}\\
    \left(\nabla_\phi\chi_\text{GPT}^2\right)_i \phi_i & = 0 \text{ for } i\in\{0,...,n-1\} \text{ (complementary slackness)}
\end{split}
\end{equation}

\paragraph{Active set approach}  A traditional active set approach starts with the assumption that all parameters (i.e. $\phi_i$) are actively constrained (set to zero).  Within this assumption, only dual feasibility can be violated, so the stepwise active set search begins by finding the worst violator of dual feasibility, which is the most negative gradient component.  The corresponding parameter is removed from the active set and added to a set of passive constraints which must have a zero gradient in order to satisfy complementary slackness.  At this point, the solution for the isolated passive set is found, which is equivalent to maximizing likelihood while considering all active parameters as constants.  Finally, primal feasibility is checked, and offending parameters are moved back to the active set.  This process is repeated while moving one parameter at a time from active to passive, until all conditions are met, or dual feasibility is met within some specified tolerance.  This approach was shown by Lawson and Hansen \cite{lawson1995solving} to converge (for qualifying optimization problems) to the correct active and passive sets in finite time.

While the original active set approach is a robust method, it is widely considered to be slow due to the need for inverting a new matrix, and performing a gradient search, for each parameter moved to the passive set.  Since Lawson and Hansen \cite{lawson1995solving}, a handful of alternative approaches to nonnegative least squares have been proposed which can improve performance in many situations \cite{chen2010nonnegativity}.  However, our needs are especially restrictive, since for the application of proton beam therapy, we would like to perform the optimization on the order of milliseconds while having $10^4$ or more parameters.  For this reason, we have taken the approach of accepting an approximate determination of the active and passive sets.  The approach is defined in Algorithm \ref{algNNGPT}.

\begin{algorithm}
\label{algNNGPT}
\SetKwInOut{Input}{input\hspace{3mm}}
\SetKwInOut{Output}{output}

\Input{$G \in \mathbb{R}^{m\times n}$, $\Sigma_p \in \mathbb{R}^{n\times n}$, $\mu \in \mathbb{R}$, $\Sigma_d \in \mathbb{R}^{m\times m}$, $d \in \mathbb{R}^m$}
\Output{$\phi$ such that $\phi \approx \argmin||d-G\phi'||^2$ subject to $\phi_i\geq 0$ for all $i$}
$P := \{0,...,n-1\}$, $\phi := \bm{0}$\;
\Repeat{$\min\phi \geq 0$}{
    $N :=$ \{$i$ where $\phi_i < 0$\}\;
    $P = P \setminus N$\;
    $\phi^N = \bm{0}$\;
    
    $\Sigma_p^P :=$ sparse $\Sigma_p$ matrix retaining only elements corresponding to $P$, but keeping the shape of $\Sigma_p$\;
    $A^P := \left[\text{sparse matrix multiply }\left(\Sigma_p^P, G^\top\right)\right]^P$\;
    $\phi^P = A^P\left[\left(\Sigma_d + \sigma_pG^PA^P\right)\setminus d\right]$\;
}
\caption{Approximate nonnegative GPT}
\end{algorithm}

In summary, Algorithm \ref{algNNGPT} starts with the assumption that all parameters (pixels) are passive (explicitly satisfying dual feasibility and complementary slackness) and steps through aggressively moving all negative pixels to the active set until both primal feasibility and complementary slackness are satisfied.  By starting with dual feasibility and working backwards, it is assumed that it is approximately satisfied once the other two sets of conditions are satisfied.  With this approach, the gradient of the $\chi^2$ function never needs to be calculated directly, and the number of steps is dramatically reduced.  We have observed that the drawbacks to the approximation can be seen at about the 3-sigma level of the reconstructions performed in section \ref{secSegmentationStudies}.  At this level, the aggressive determination of actively-constrained pixels can be seen as a noticeably harsh outline of the fitted distribution.

\subsection{Diffusion modeling}
\label{subsecDiffusion}
When measuring the distribution of charge extracted at the electrodes of a detector, one measures the distribution of power deposited into the detector convolved with diffusion of charge carriers as they propagate towards the electrodes.  Our construction of $G$ (equation~(\ref{eqDesignMatrix})) effectively deconvolves the effects of diffusion as part of the regularized $\chi^2$ fit by including a model for the diffusion.  This model can be chosen depending on the particular detector configuration and requirements of the reconstruction.  A description of the model used herein and the benefits of tuning the diffusion are given below.


\paragraph{Diffusion model} Theoretically, the transverse diffusion in a parallel plate detector is given by a normal distribution with $\sigma=\sqrt{\frac{2\epsilon_k}{eE}h}$ where $\epsilon_k$ is the characteristic energy, $e$ is the elementary charge, and $E$ is the electric field strength in the ionization chamber \cite{Palladino1975drift}.  Herein we model the diffusion to be normally distributed for a given height above a sensitive electrode plane with standard deviation proportional to the square root of the initial height, $h$. There is some max height (the distance between electrode planes, labeled "sensitive region" in Figure \ref{figSegmentation}), such that the diffusion attains maximum standard deviation $\sigma_\text{max}$.  We can parameterize the diffusion to have standard deviation $\sigma_\text{max}\sqrt{\eta}$, where $\eta \in (0, 1)$ represents the relative initial height.  For each sample used during Monte Carlo generation of the design matrix, values of $\eta$, $x$, and $y$ are chosen according to equation~(\ref{eqDiffModel}) and $x$ and $y$ are added to the coordinates chosen from the geometry of the pixel.
\begin{equation}
\label{eqDiffModel}
\begin{split}
    \eta & \sim \mathcal{U}\left(0, 1\right) \\
    x,y & \sim \mathcal{N}\left(0,\sigma_\text{max}\sqrt{\eta}\right)
\end{split}
\end{equation}


\begin{figure}
    \centering
    \includegraphics[width=5.25in]{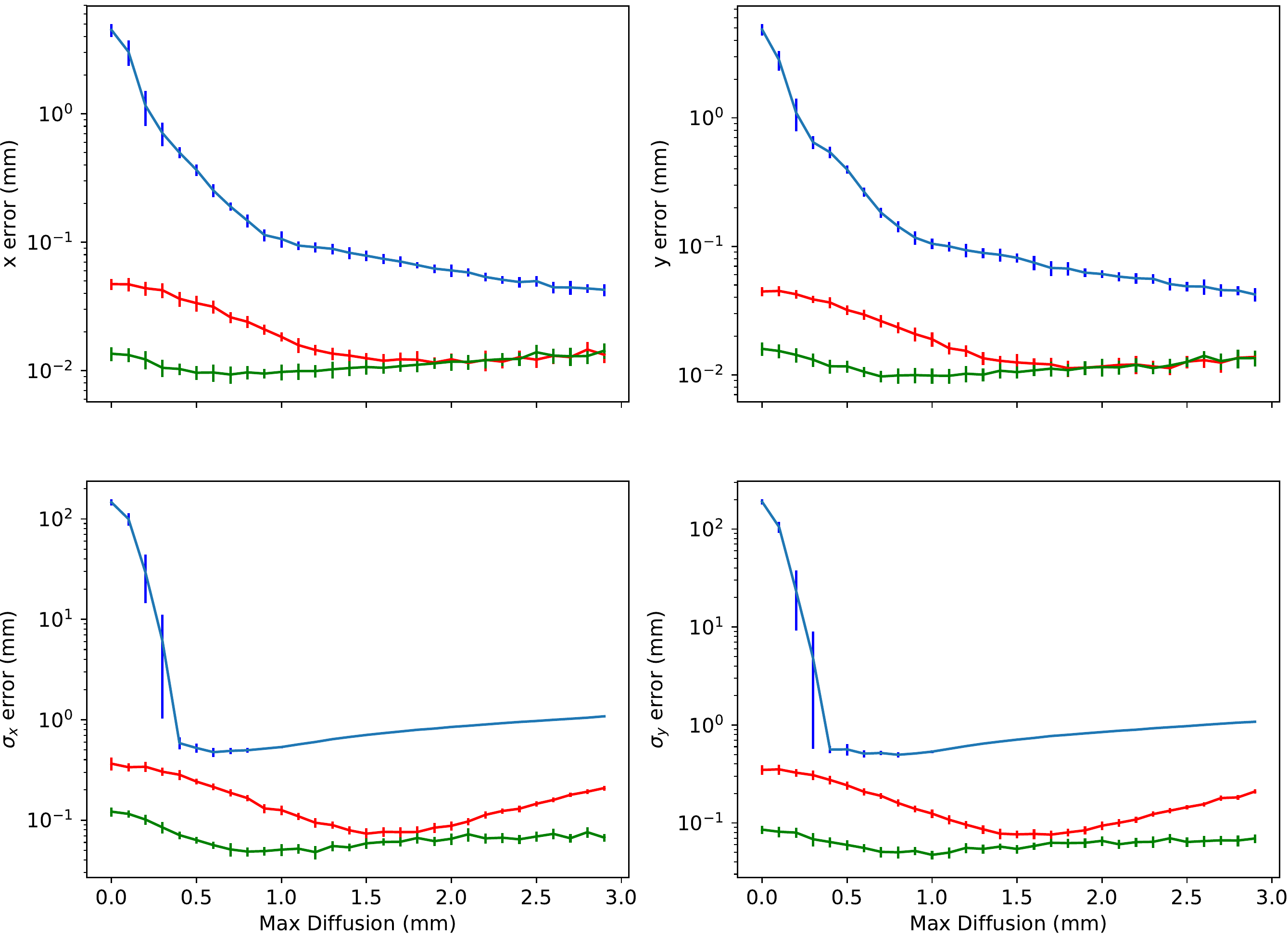}
    \caption{Shown here is the accuracy of reconstruction with a particular 3-D segmentation scheme for delta function beams in blue, half-pitch sigma beams in red (where pitch refers to the spacing between lines in a projection) and full-pitch sigma beams in green with varying diffusion. On the top is the error in the mean position in both x (top-left) and y (top-right). On the bottom is the error in the standard deviation in both x (bottom-left) and y (bottom-right). In each plot the x-axis is $\sigma_\text{max}$ from equation~(\ref{eqDiffModel}).}
    \label{fig:charge_sharing}
\end{figure}

\paragraph{Improvements due to charge sharing} Detectors are often designed to minimize diffusion in order to avoid the need for deconvolution. While this successfully minimizes error within the assumption of zero diffusion, highly localized signals may be difficult to reconstruct due to a lack of charge sharing between neighboring readout channels.  Deconvolving the effects of diffusion as described above allows for flexibility in the amount of diffusion, so that significant amounts of diffusion can be used to spread out localized signals.
The charge sharing that is enabled by the diffusion greatly improves the reconstruction accuracy for localized signals.

The accuracy of reconstructing a pencil beam with various widths is shown in Figure~\ref{fig:charge_sharing} as a function of maximum diffusion, $\sigma_\text{max}$. While this behavior is specific to the 3-D readout segmentation shown in Figure \ref{figExampleReadout} as well as the diffusion modeled by equation \ref{eqDiffModel}, it gives an idea of the sort of improvements in accuracy that can be made through charge sharing. The figure shows that the error in the mean position improves monotonically with increasing diffusion for very narrow beams, while the error in the standard deviation reaches a minimum in each case after which it increases slightly. In the case of a delta function beam, the beam cannot be localized without significant diffusion.
Even when the beam sigma is equal to the projection pitch there are slight improvements in the accuracy of the reconstruction of the spread when diffusion is included.

\section{Segmentation studies}
\label{secSegmentationStudies}
\begin{figure}
    \centering
    \includegraphics[width=1.25in]{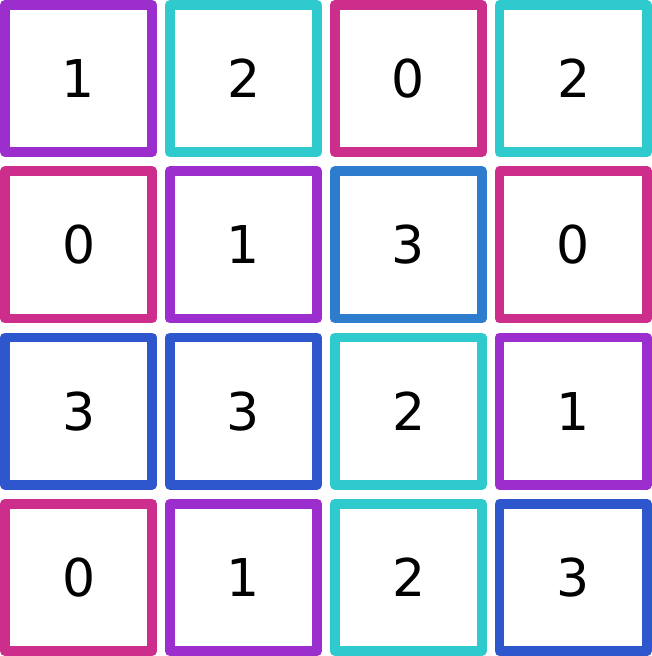}
    \hspace{0.25in}
    \includegraphics[width=1.25in]{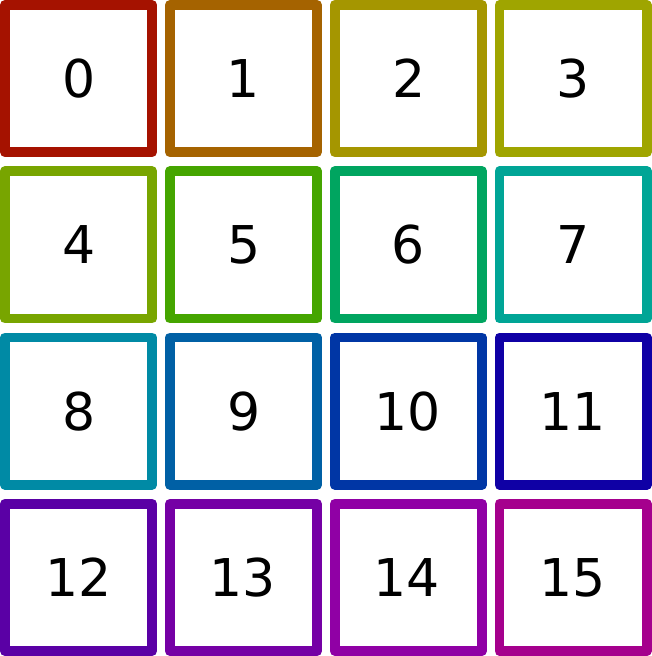}
    \caption{\textbf{Example coded and pixelated electrode segmentation} - Shown are examples of coded (left) and pixelated (right) segmentation schemes.  Notice the random, repeated channel labeling in the coded scheme, and the unique channel association in the pixelated scheme.}
    \label{figCodedPixelated}
\end{figure}

So far we have intentionally kept the concept of detector electrode segmentation highly generic.  In this section, we apply the formulation from section~\ref{secGPT} to a handful of interesting particular cases in order to qualify the potential for generality in such a formulation, as well as to motivate novel segmentations, especially within the context of proton therapy pencil beam scanning.  The cases utilize single-plane segmentations which we refer to as 2-D, 3-D, coded, and pixelated.  We use the term coded segmentation by analogy with coded apertures used in x-ray and gamma-ray imaging \cite{finger1985hexagonal}.  In our case, coded effectively means that for a regular pattern of electrode pads, the pads are randomly assigned a channel from a pool of channels which are reused, and we completely depart from the concept of linear projections.  The term pixelated refers to a regular pattern of pads, where there is a one-to-one mapping between pad and readout channel.  While Figure~\ref{figExampleReadout} showed example representations of 2-D and 3-D segmentation, Figure~\ref{figCodedPixelated} shows example representations of coded and pixelated segmentation.

\begin{figure}
    \centering
    \includegraphics[width=2in]{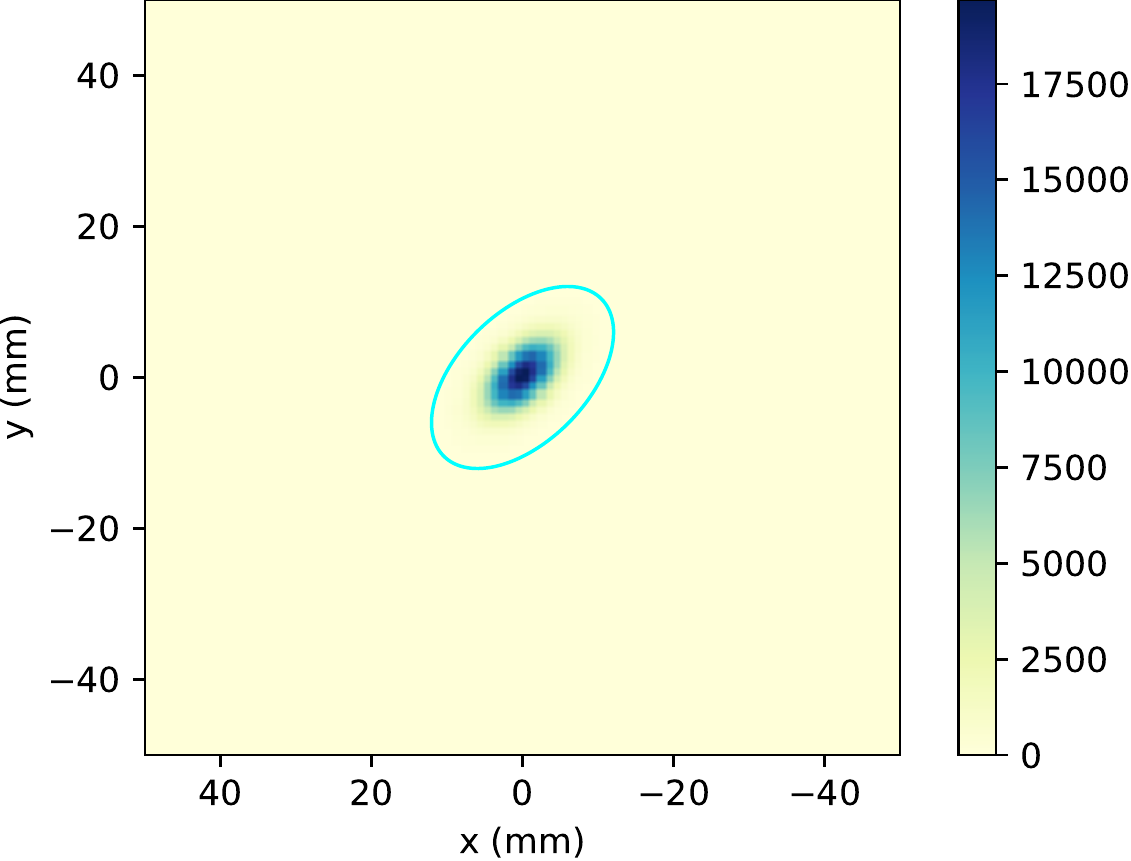}\hspace{0.25in}
    \includegraphics[width=2in]{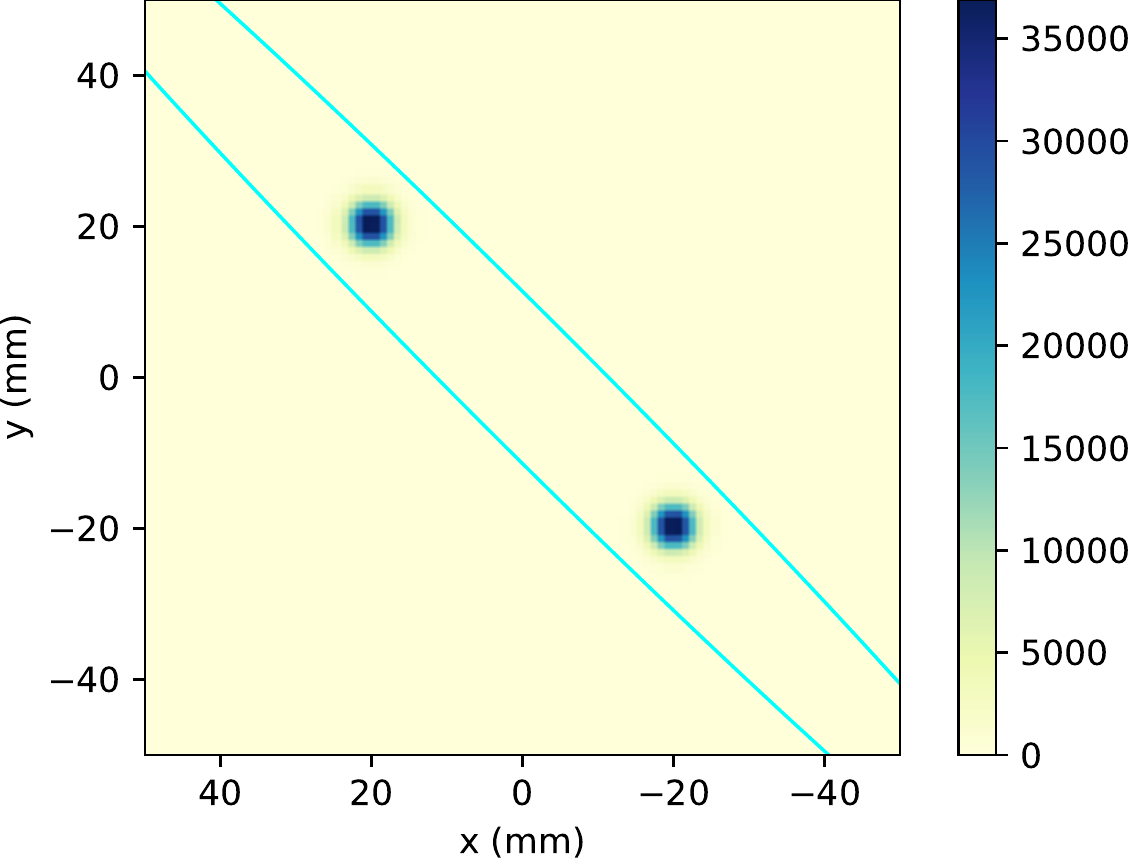}
    \\\vspace{0.25in}
    \includegraphics[width=2in]{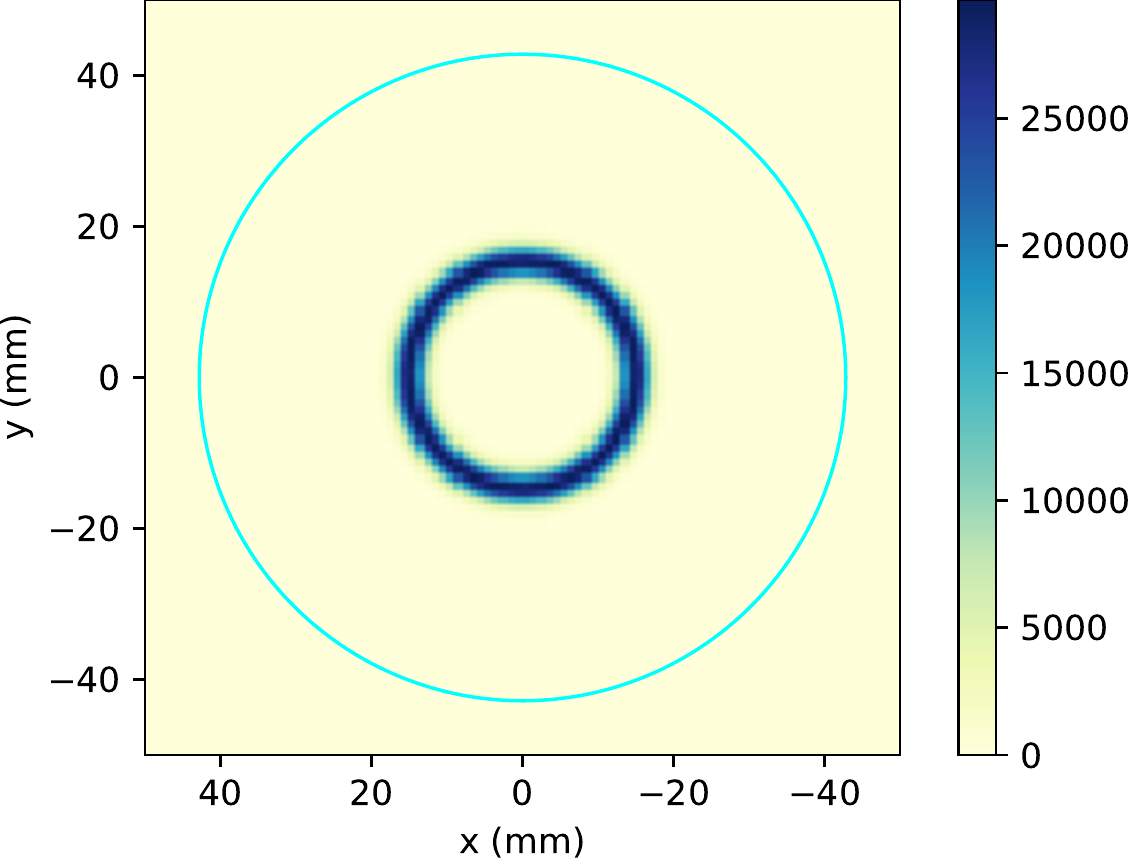}\hspace{0.25in}
    \includegraphics[width=2in]{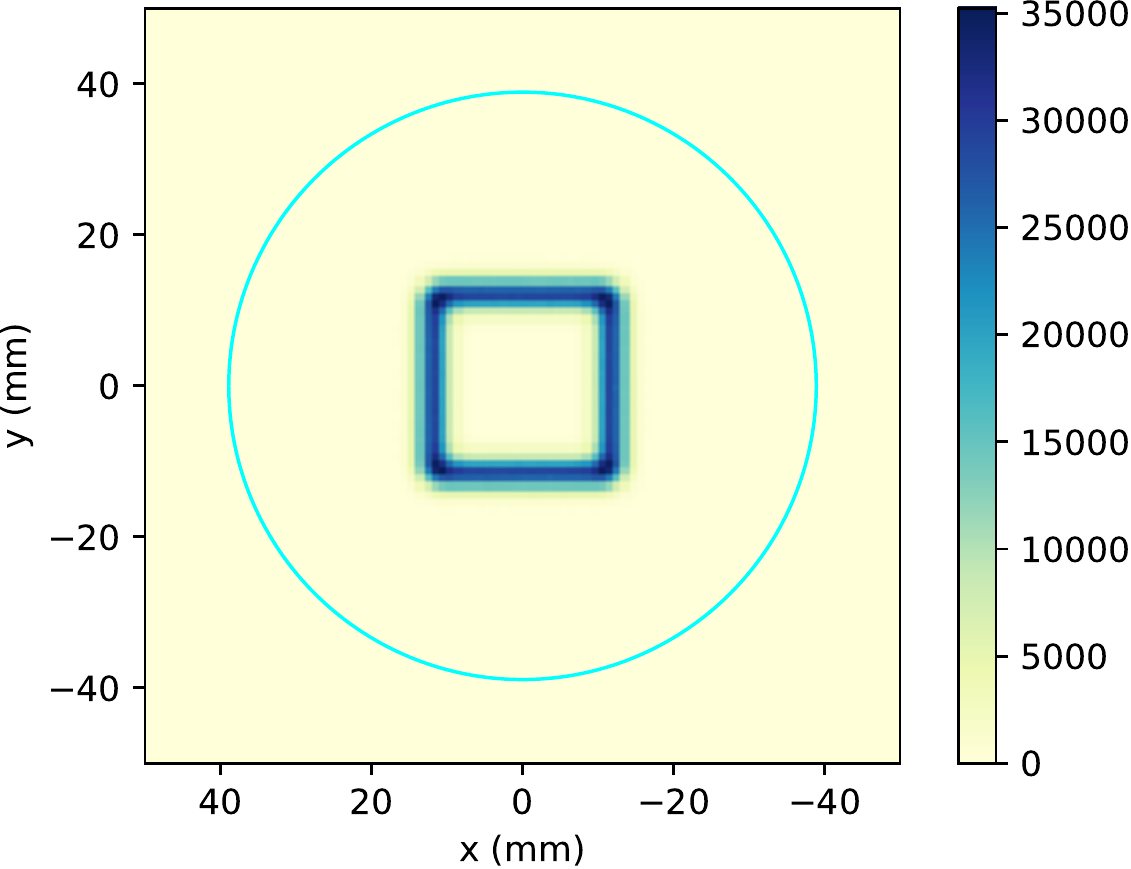}
    \caption{\textbf{Truth images for benchmark simulated beam profiles} - Each of these images represents a simulated distribution of energy deposition into a detector used to benchmark the imaging capabilities of various segmentation schemes.  Cyan lines show 4-sigma levels to indicate the second mean moments of the distributions in pixel coordinate space.  Top left: covariant Gaussian with diagonal covariance elements of 9~mm$^2$ and off diagonal elements of -4.5~mm$^2$.  Top right: dual Gaussians with $\sigma_x,\sigma_y=2$~mm.  Bottom left: circle of radius 15~mm convolved with 2~mm Gaussian.  Bottom right: square with perimeter equal to the circle case, convolved with 2~mm Gaussian.}
    \label{figBenchmarkProfiles}
\end{figure}

For each case, the formulation in section \ref{secGPT} is applied to a simulated detector with a set of benchmark simulated beam profiles.  The detector in each case covers an area of 100~cm$^2$ and utilizes exactly 180 readout channels, and the reconstructed images have 1~mm pixel pitch and a resolution of 100x100 ($10^4$ total virtual pixels).  Monte Carlo samples for each benchmark beam profile are binned into the channel space for each segmentation case ($d$ from equation~(\ref{eqLinearRelationship})), and Algorithm \ref{algNNGPT} is applied.  The characteristic length $l$ (equation~(\ref{eqSEKernel})) is set to 1~mm, the kernel sparsity cutoff (equation~(\ref{eqSEKernelCutoff})) is set to 0.1, and the maximum diffusion standard deviation (equation~(\ref{eqDiffModel})) is set to 2~mm.  The four benchmark beam profiles used are shown by their truth images in Figure~\ref{figBenchmarkProfiles}.  The exact code used to generate the figures presented below can be found in \cite{nngptRepo}.

\begin{figure}[p]
    \centering
    \includegraphics[width=2in]{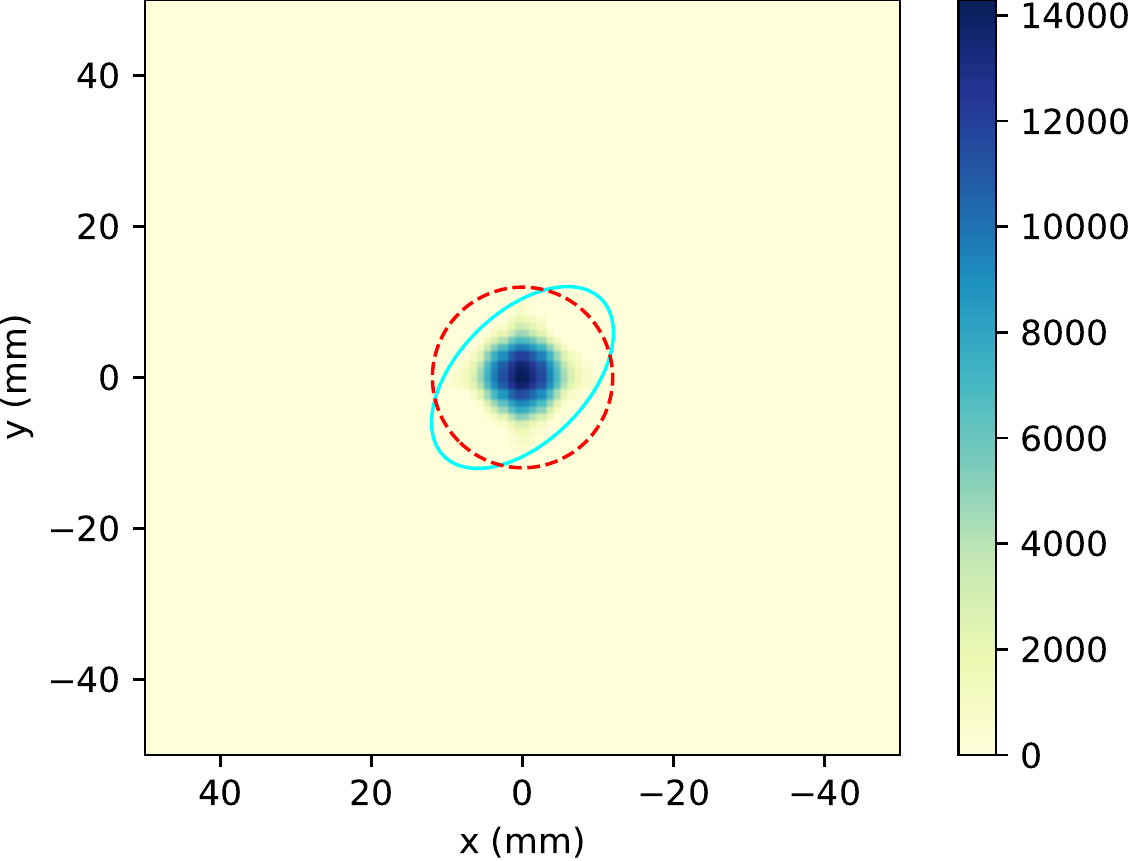}\hspace{0.25in}
    \includegraphics[width=2in]{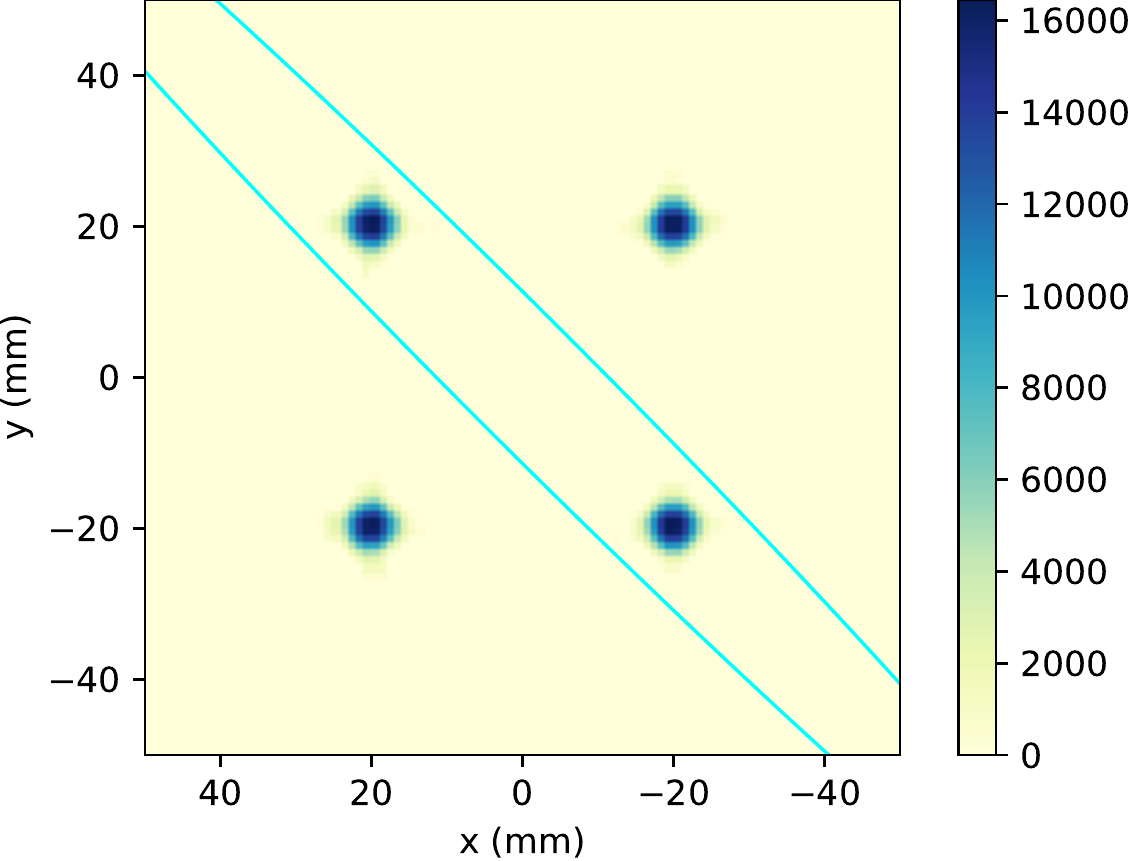}
    \\\vspace{0.25in}
    \includegraphics[width=2in]{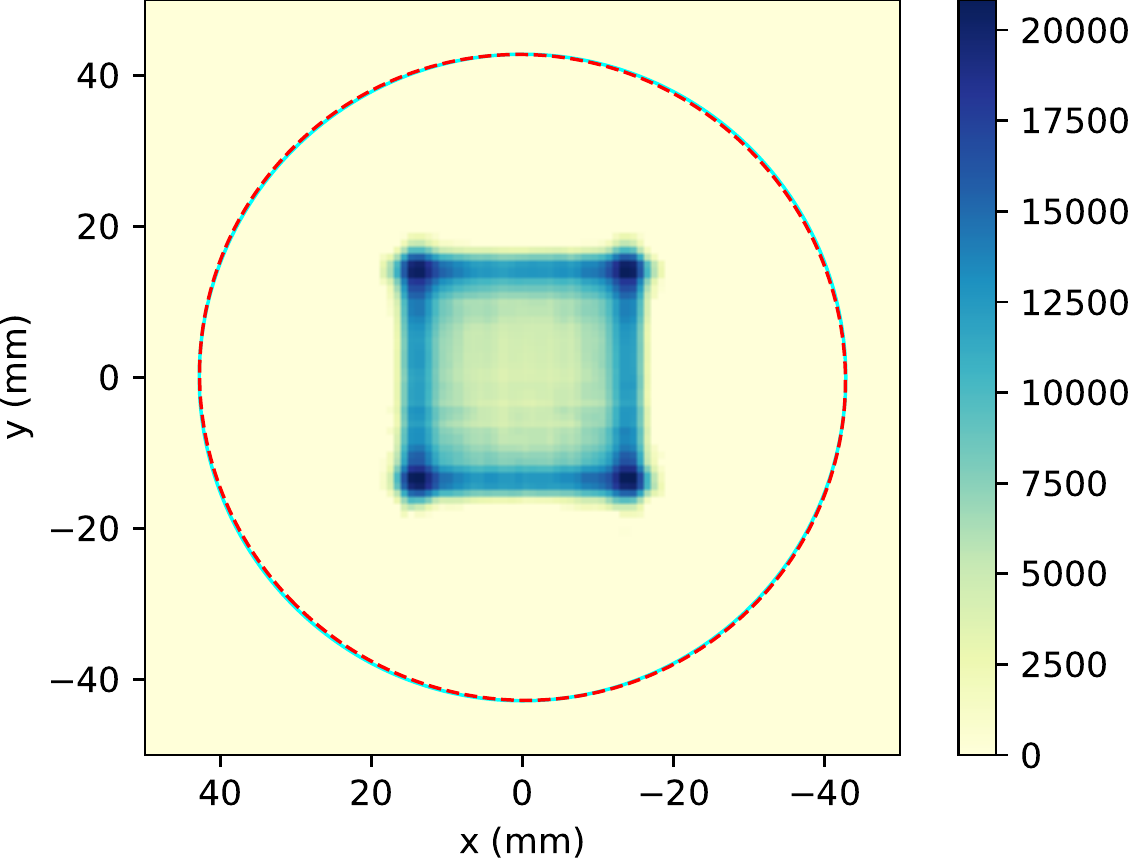}\hspace{0.25in}
    \includegraphics[width=2in]{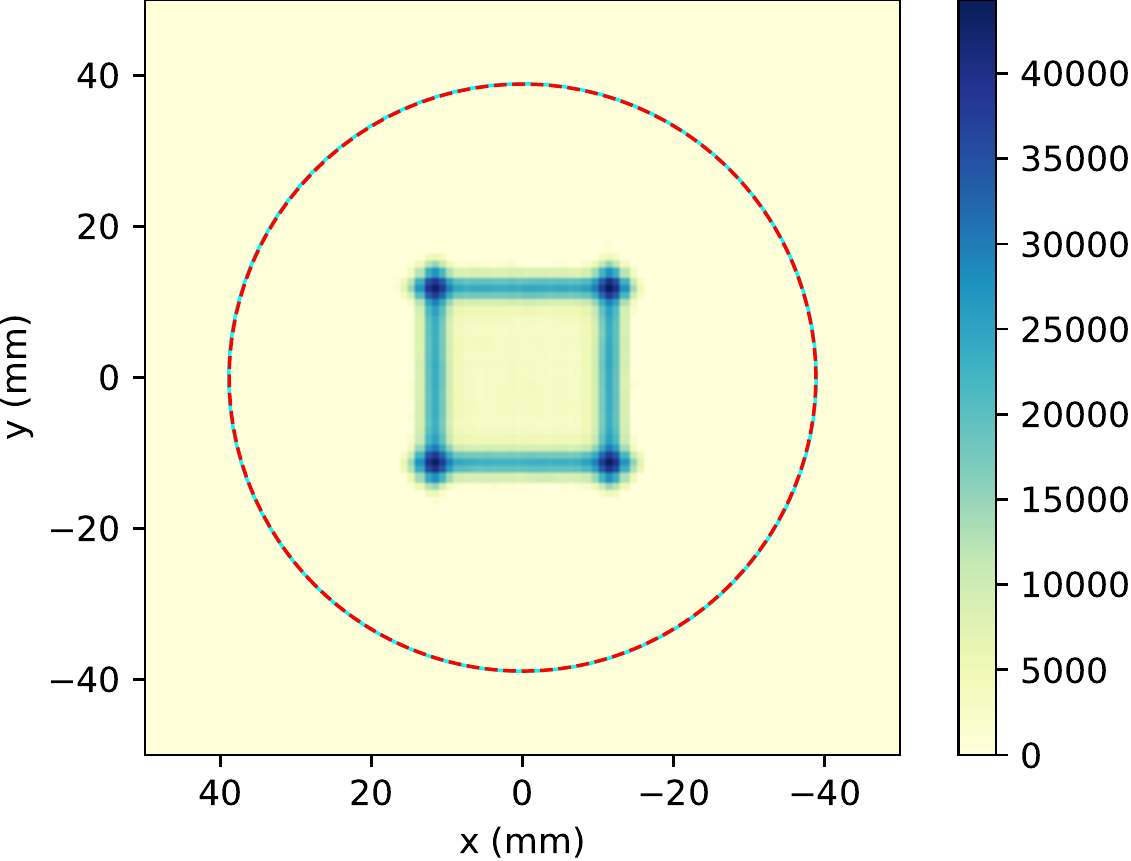}
    \caption{\textbf{Image reconstruction with 2-D projection segmentation} - Top left: covariant Gaussian, top right: double Gaussian, bottom left: circle, bottom right: square.  Cyan lines show the truth 4-sigma levels, and red lines indicate reconstructed 4-sigma levels.}
    \label{fig2DTomo}
\end{figure}

\paragraph{2-D projection segmentation}  For a 2-D projection case, we use a 90x90 square grid of pads for a total of 180 channels between the x and y projections.  Figure \ref{fig2DTomo} shows the results of reconstructing the benchmark distributions.  The reconstruction demonstrates that such a 2-D projection cannot provide significant information about the covariance of a distribution, and artifacts are dominant with any sort of complexity.  The double Gaussian case demonstrates that the tomography formulation will split the reconstructed image between degenerate possibilities.

\begin{figure}[p]
    \centering
    \includegraphics[width=2in]{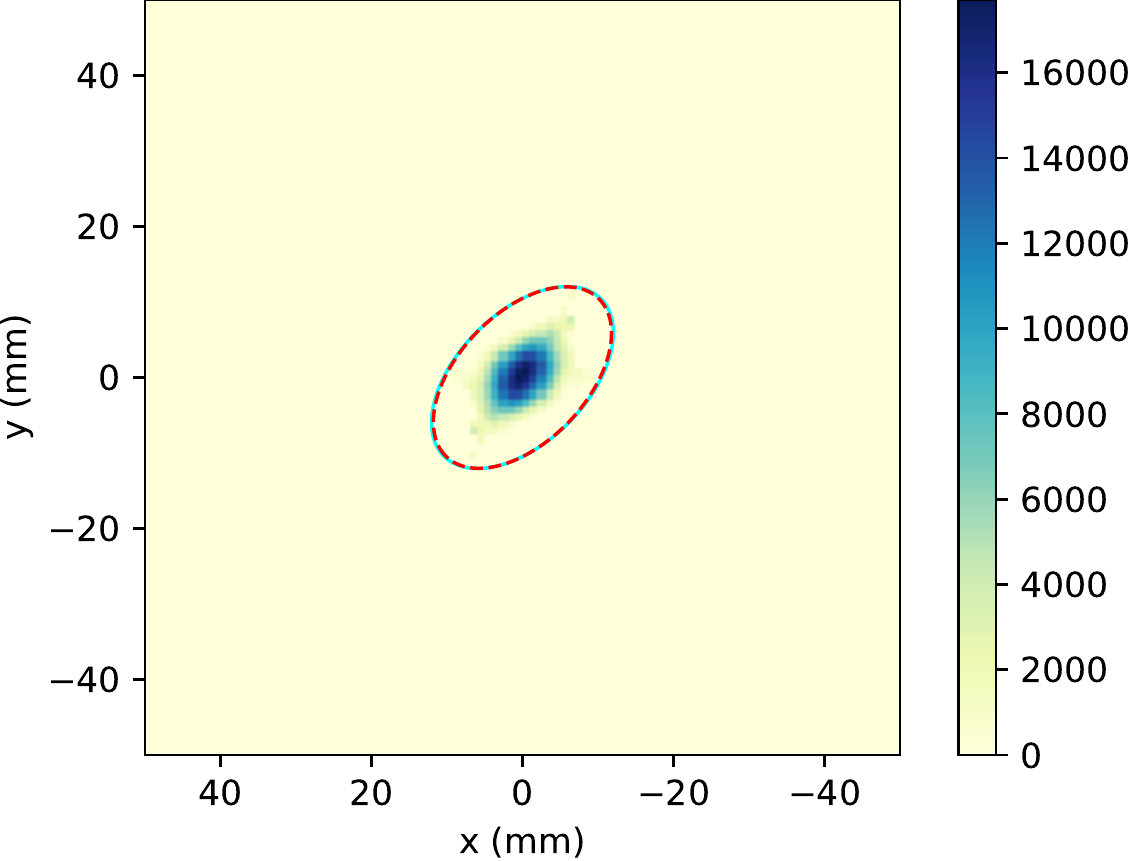}\hspace{0.25in}
    \includegraphics[width=2in]{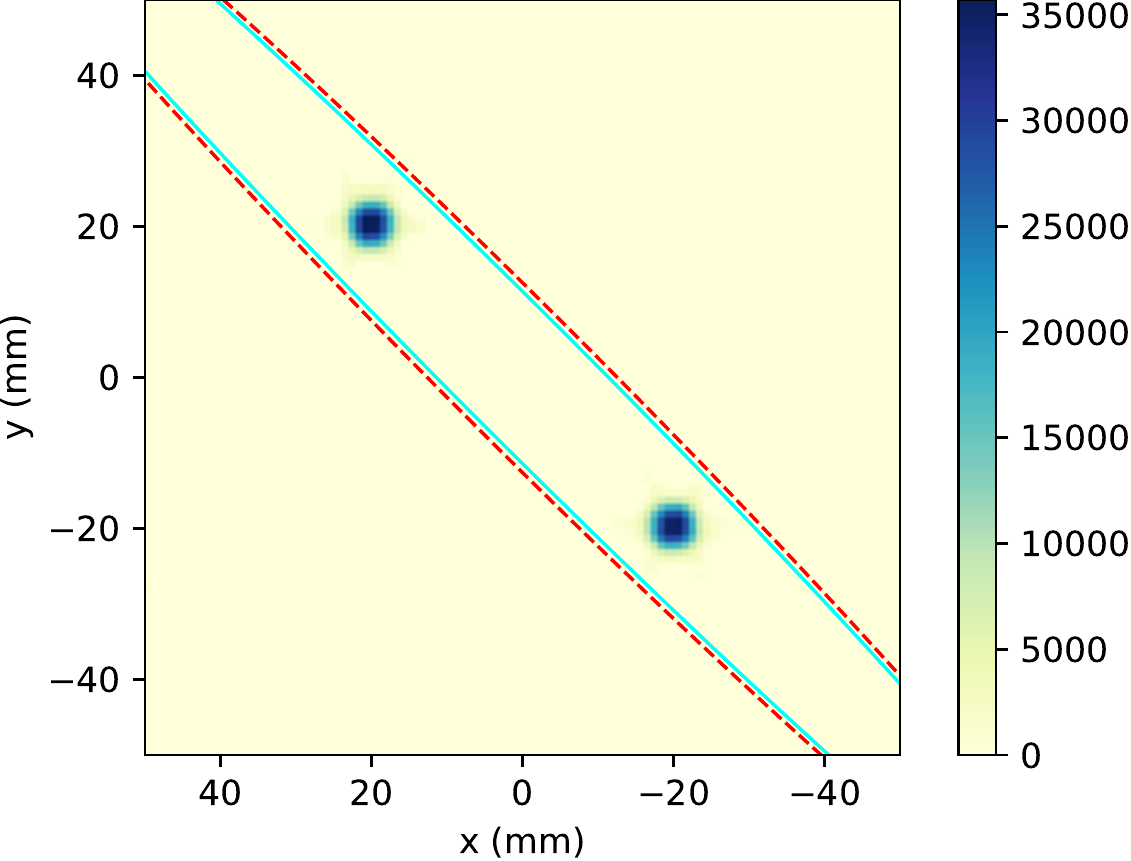}
    \\\vspace{0.25in}
    \includegraphics[width=2in]{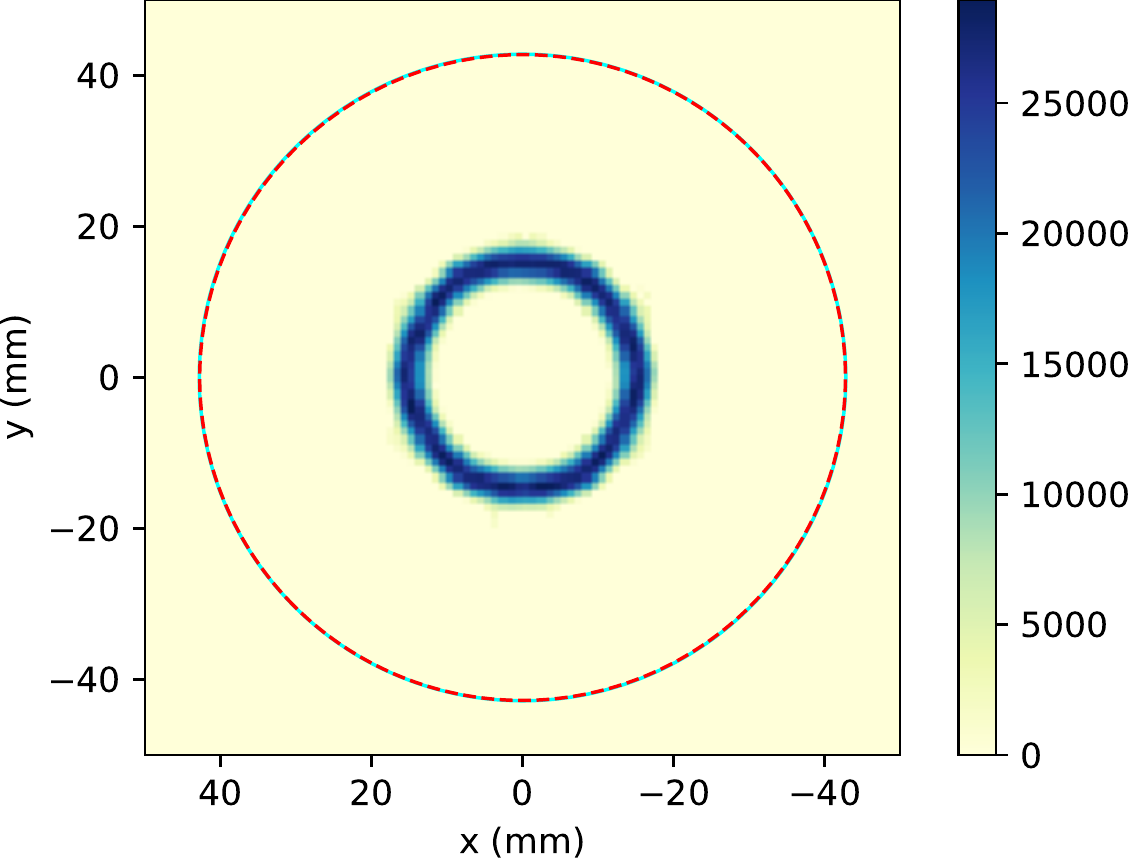}\hspace{0.25in}
    \includegraphics[width=2in]{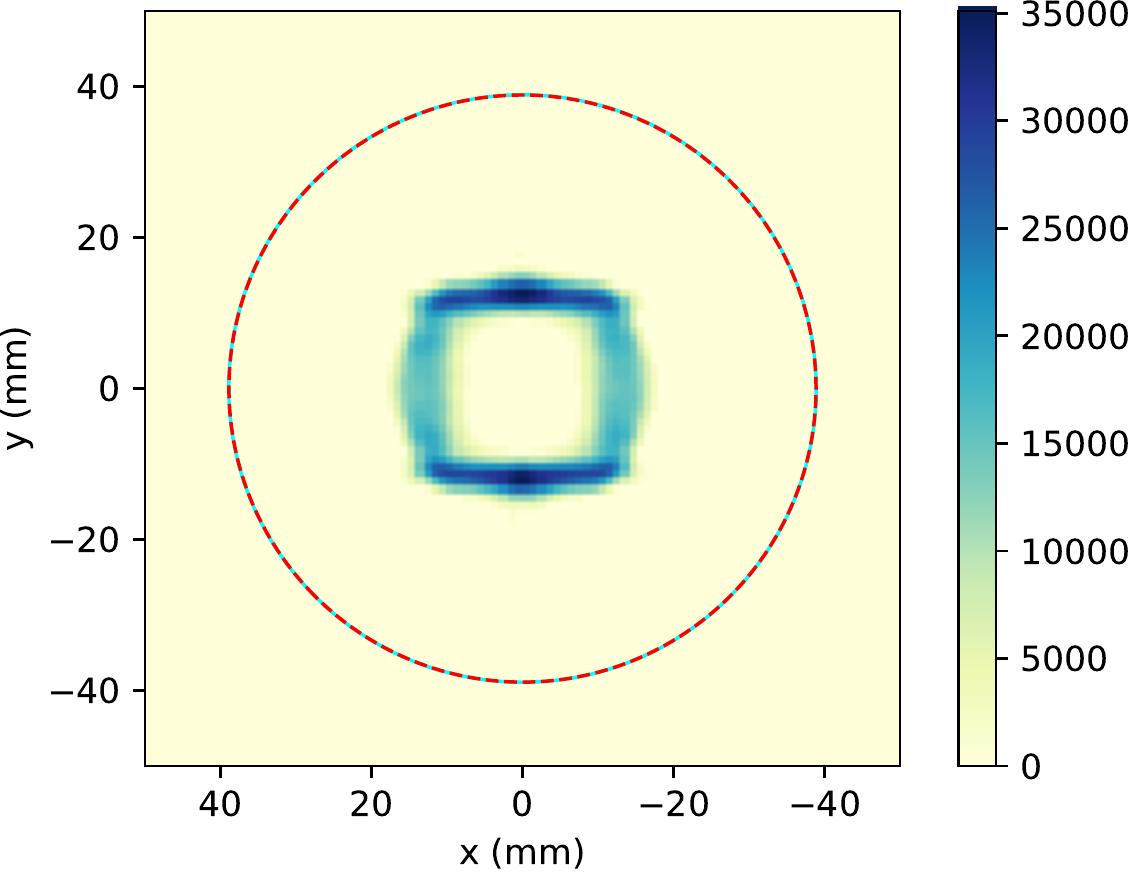}
    \caption{\textbf{Image reconstruction with 3-D projection segmentation} - Top left: covariant Gaussian, top right: double Gaussian, bottom left: circle, bottom right: square.  Cyan lines show the truth 4-sigma levels, and red lines indicate reconstructed 4-sigma levels.}
    \label{fig3DTomo}
\end{figure}

\paragraph{3-D projection segmentation}  For a 3-D projection case, we use a hexagonal grid of pads (see Figure \ref{figExampleReadout}) with 60 channels for each projection and a total sensitive area that matches the 2-D case.  Figure \ref{fig3DTomo} shows the results of reconstructing the benchmark distributions.  We see in the figure that the 3-D projection can precisely reconstruct arbitrary covariance in the distribution.  The discrete symmetry of the projections is not nearly as evident as in the 2-D case, though it can still be seen.  The square benchmark emphasizes the limitations of such a readout for arbitrary distributions.

\begin{figure}[p]
    \centering
    \includegraphics[width=2in]{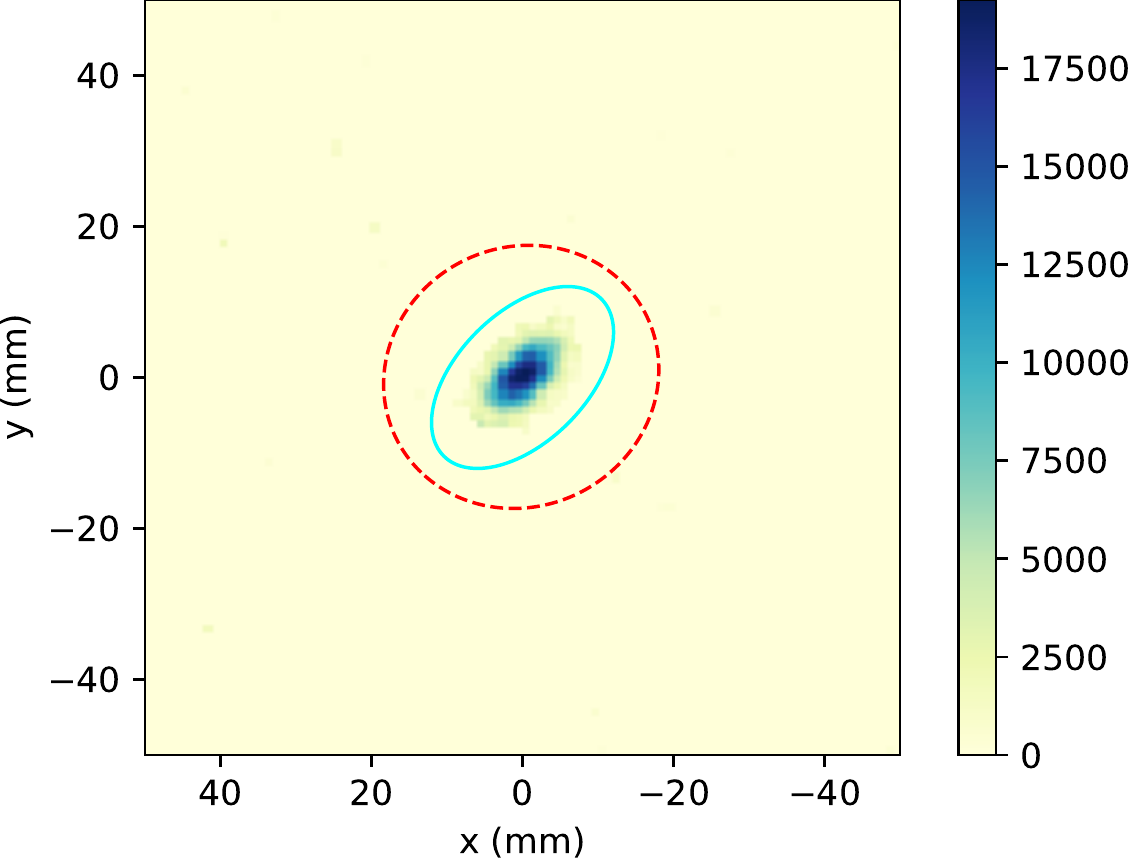}\hspace{0.25in}
    \includegraphics[width=2in]{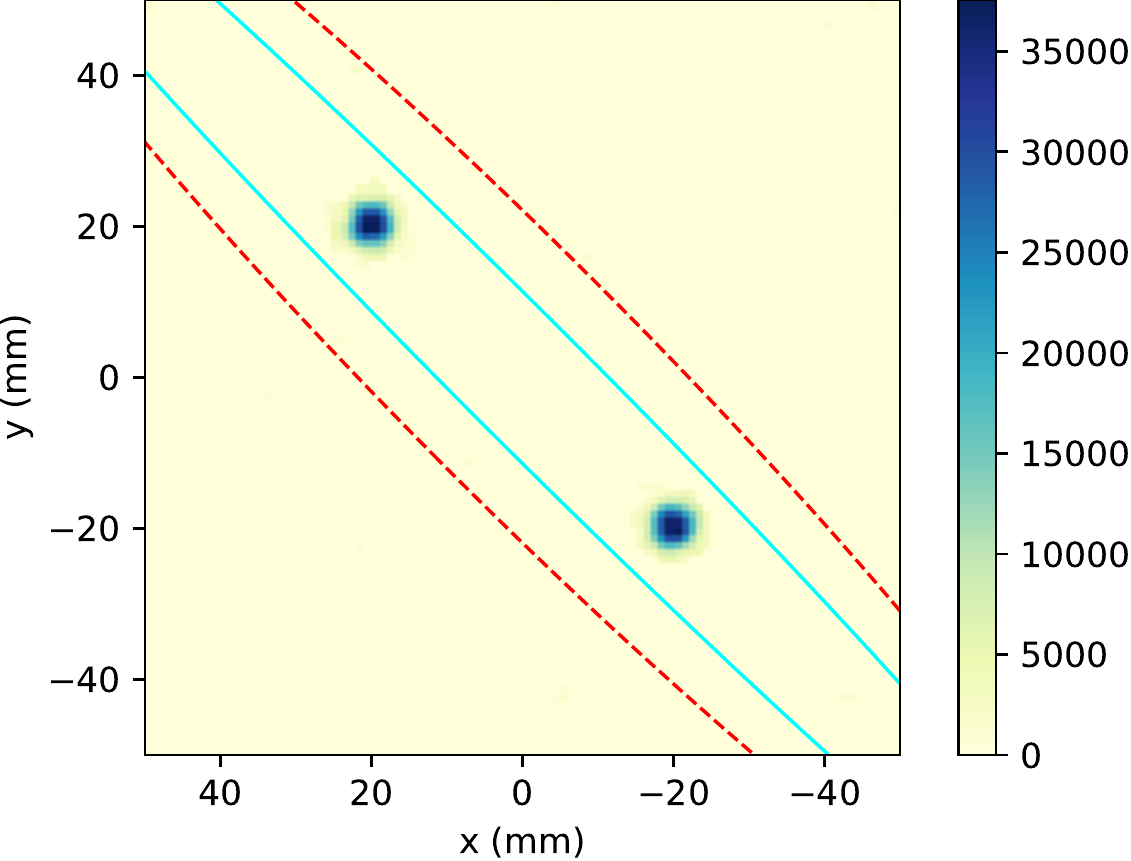}
    \\\vspace{0.25in}
    \includegraphics[width=2in]{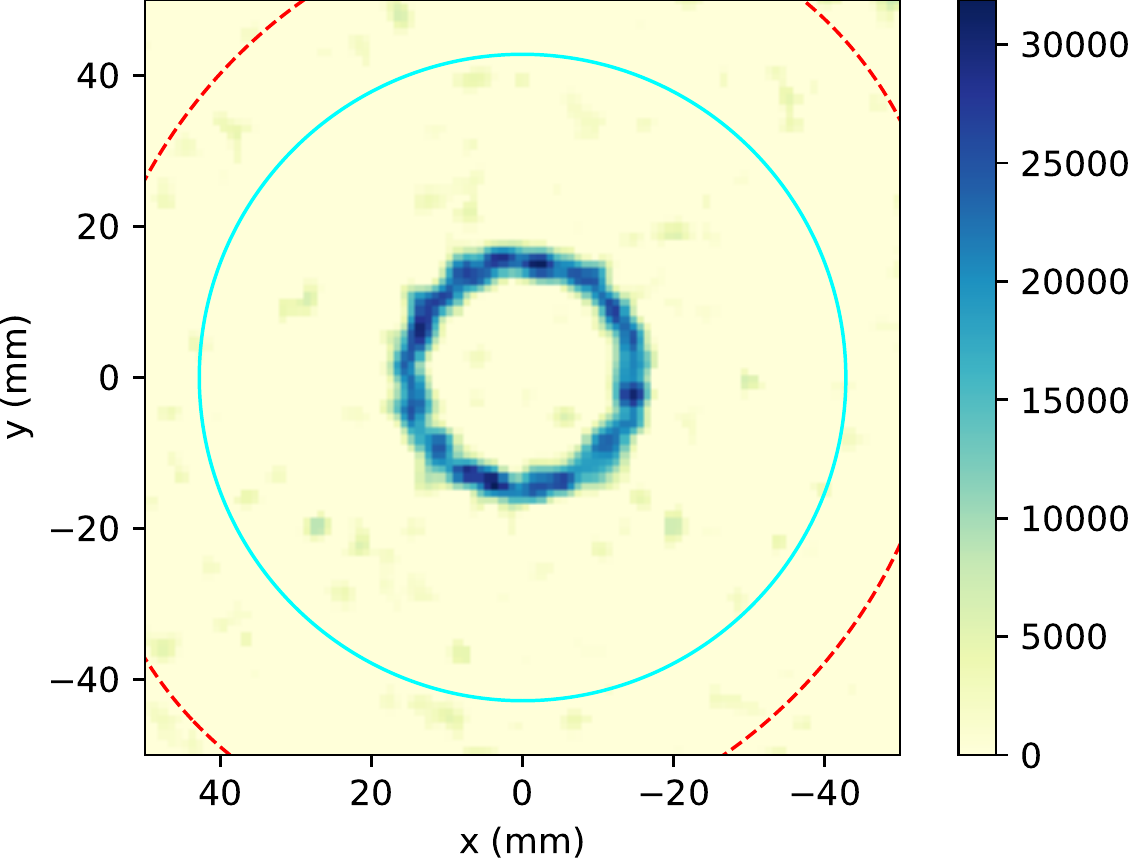}\hspace{0.25in}
    \includegraphics[width=2in]{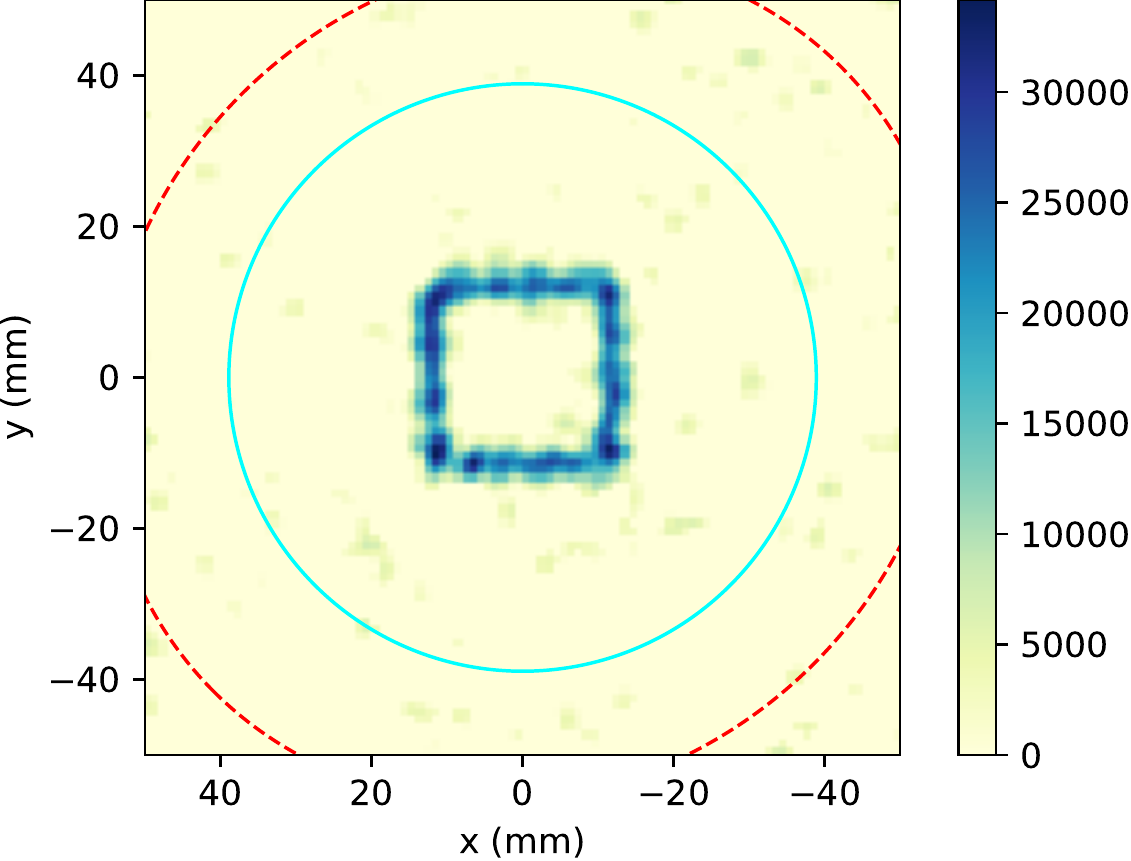}
    \caption{\textbf{Image reconstruction with coded segmentation} - Top left: covariant Gaussian, top right: double Gaussian, bottom left: circle, bottom right: square.  Cyan lines show the truth 4-sigma levels, and red lines indicate reconstructed 4-sigma levels.}
    \label{figCodedTomo}
\end{figure}

\paragraph{Coded segmentation}  For a coded segmentation case, we use a square grid of 100x100 pads that are each randomly connected to one of 180 channels.  Figure \ref{figCodedTomo} shows the results of reconstructing the benchmark distributions.  In this case we demonstrate generalizing the concept beyond linear projections, and find that while discrete symmetries do not appear, the ability to characterize the shape and location of the distribution is significantly degraded.  Additionally, the noise seen throughout the image grows as the size of the distribution (portion of detector with significant signal) increases.  This can be counteracted by adding more readout channels, but the comparison given here purposefully fixes the channel count to 180.  It should be noted that no effort was made to disallow nearby virtual pixels from being assigned the same channel.  The channel assignment scheme was kept intentionally simple, but we believe that more complex schemes can achieve better performance.

\begin{figure}[p]
    \centering
    \includegraphics[width=2in]{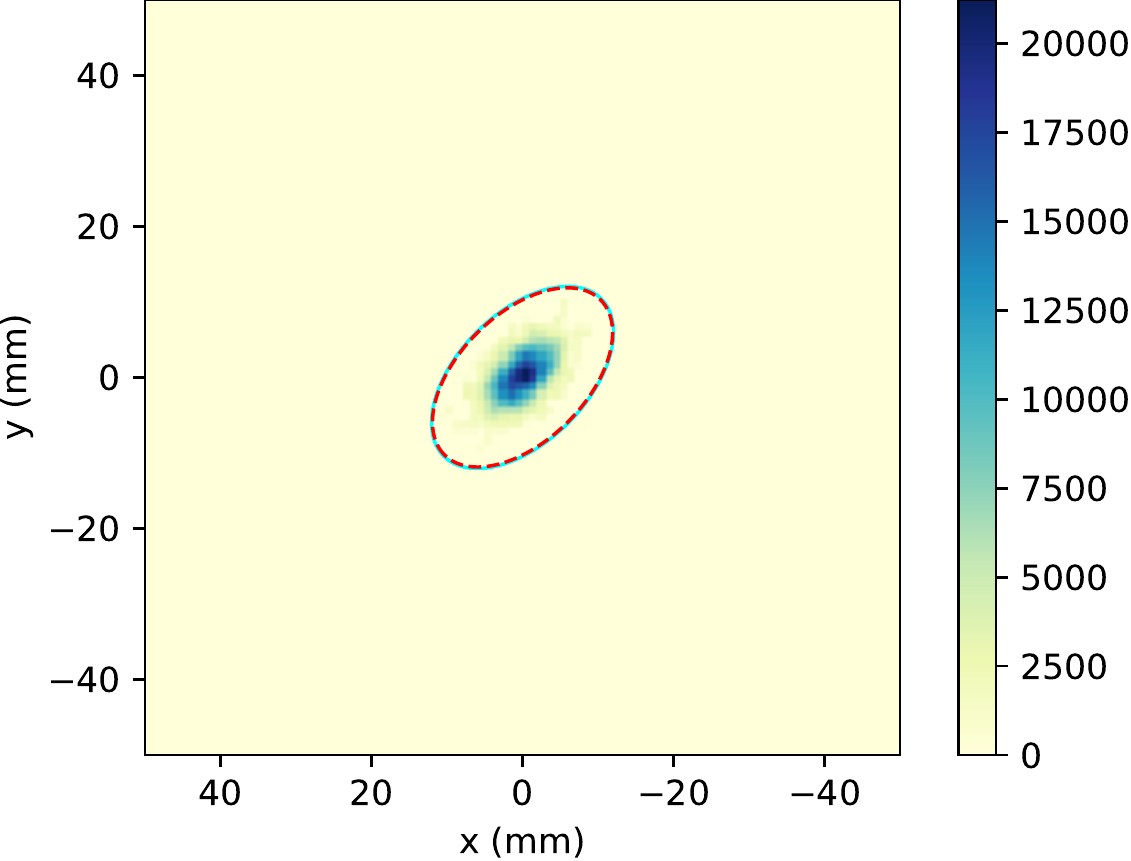}\hspace{0.25in}
    \includegraphics[width=2in]{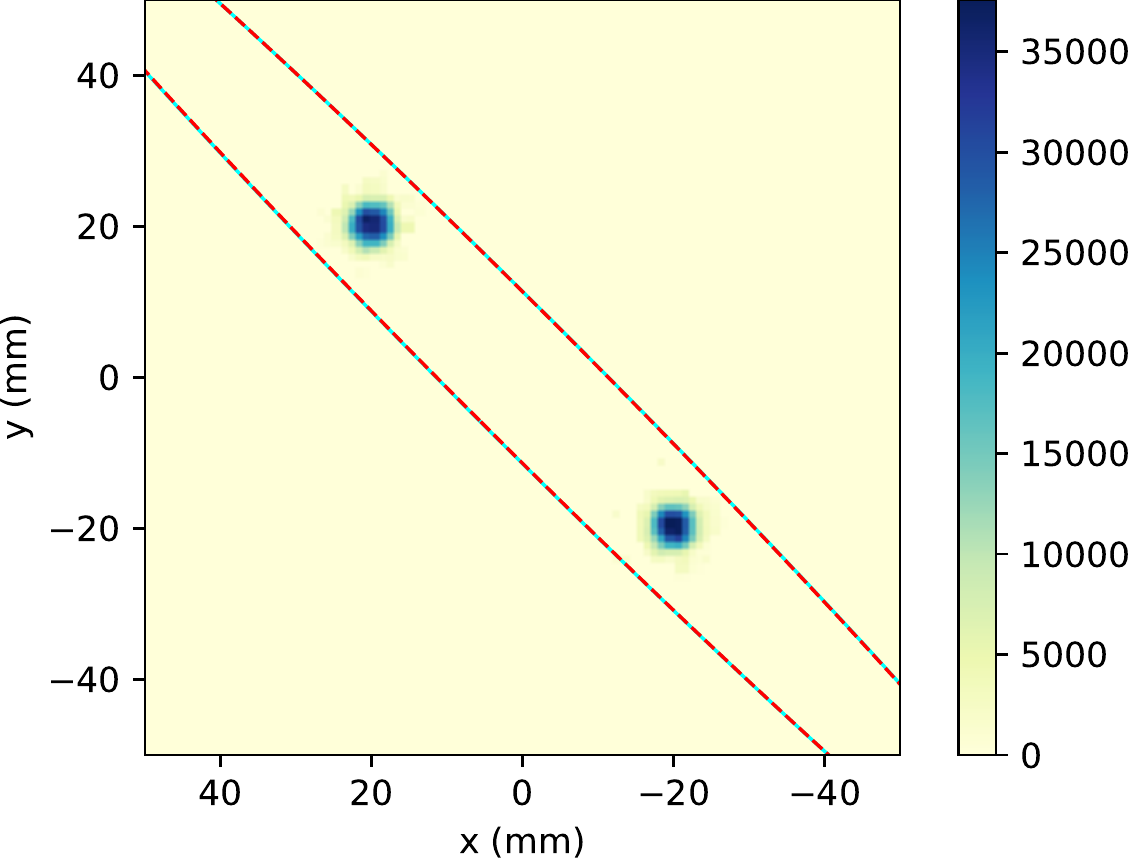}
    \\\vspace{0.25in}
    \includegraphics[width=2in]{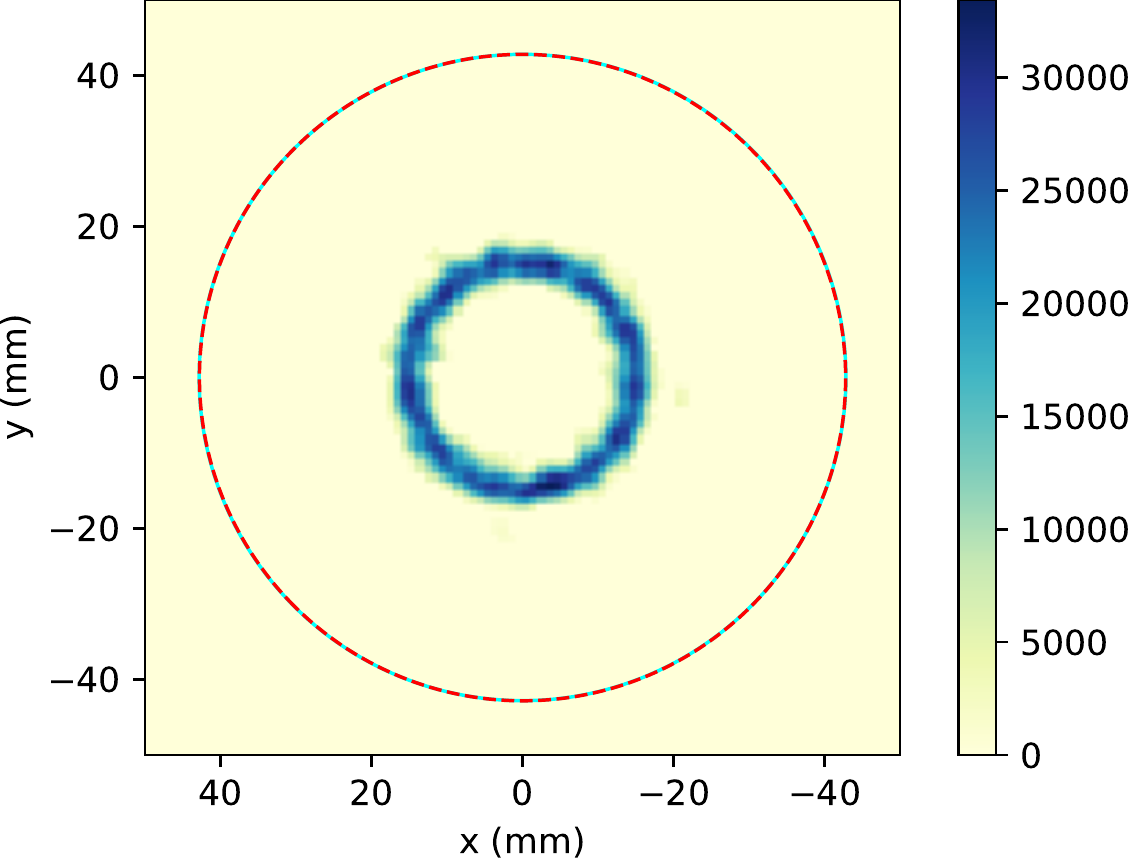}\hspace{0.25in}
    \includegraphics[width=2in]{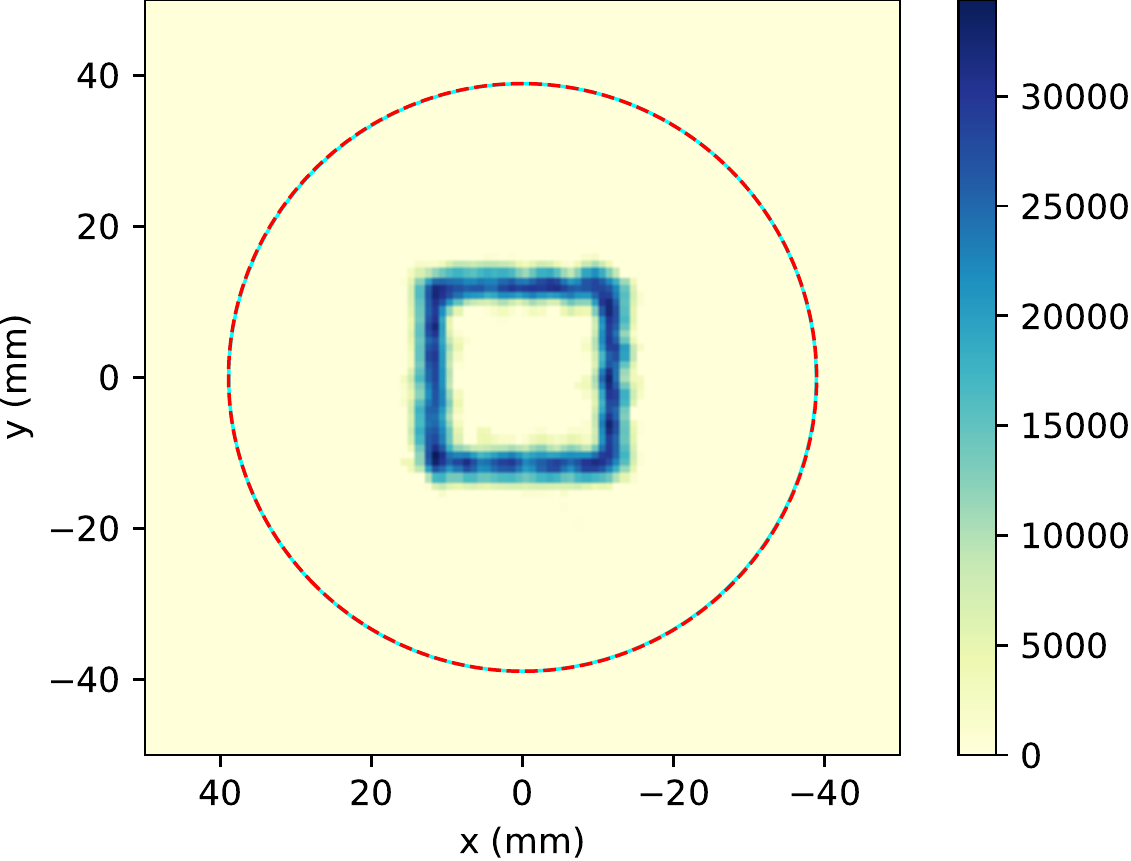}
    \caption{\textbf{Image reconstruction with dual-plane coded/coarsely-pixelated segmentation} - Top left: covariant Gaussian, top right: double Gaussian, bottom left: circle, bottom right: square.  Cyan lines show the truth 4-sigma levels, and red lines indicate reconstructed 4-sigma levels.}
    \label{figCodedPixelatedTomo}
\end{figure}

\paragraph{Hybrid coded and pixelated segmentation}  In this case, we test the additional concept of instrumenting both the anode and the cathode of the detector.  The purpose here is to augment the imaging capabilities of the coded segmentation with information that can better localize the energy deposition.  We chose a 100x100 square grid of pads that are each randomly connected to one of 99 channels on one side of the detector, and a 9x9 square grid of pads that are uniquely mapped to an additional 81 channels on the other side of the detector.  Figure \ref{figCodedPixelatedTomo} shows the results of reconstructing the benchmark distributions.  In this case, we see very precise reconstruction of the mean position and covariance of the distributions, and reasonable imaging capabilities.

\section{Discussion and Conclusions}
We have motivated and explored the concept of generalizing strip-like electrode segmentations in planar detectors using an approach based on Gaussian process tomography.  For sparse detector depositions, e.g. narrow beams, we find that this approach can effectively improve detection resolution compared to a standard pixelated segmentation for a given number of readout channels.  In section \ref{secSegmentationStudies} we studied simulated configurations with 180 readout channels over a 100~cm$^2$ area.  In a standard pixelated scenario, this channel count and sensitive area would have resulted in a pixel pitch of 7.5~mm, but in our simulated non-pixelated cases we resolve features that are significantly smaller.  Four different segmentation schemes were studied for the purpose of comparing and contrasting the imaging information provided by each.  In a practical application, the appropriate kind of segmentation would be chosen, and is not necessarily one of the cases we have presented.  In section \ref{secGPT} we laid out a generic approach for any kind of segmentation, and future studies can build off of our approach with specialized segmentations.

A critical part of the image reconstruction presented here is the enforcement of nonnegativity of the virtual pixels.  This introduced a significant cost in CPU time.  However, an approximate approach was used in order to keep calculation times reasonable, particularly for use in the context of scanning proton beam therapy.  For the reconstructed images presented (with $10^4$ virtual pixels and 180 channels), the average calculation time on a multi-core AMD Ryzen 5 CPU was 60-70~ms depending on the benchmark distribution, and $\sim30$~ms on GPUs attached to Google Cloud Platform instances.  In order to achieve this, an optimized Tensorflow graph was constructed (see \cite{nngptRepo} for code).

Physical implementations of the presented segmentation cases are relatively trivial in some cases, and highly complex in others.  The 2-D case can use simple trace routing on a PCB, for example.  The 3-D case has been built and tested by Radiation Detection and Imaging, LLC in collaboration with Arizona State University for use in proton beam therapy quality assurance.  The coded cases in general pose a significant challenge in trace route optimization in a PCB implementation.  At the moment, the coded cases are an academic excursion, but we believe that physical implementations of similar schemes are plausible.

\paragraph{Acknowledgement}  This material is based upon work supported by the U.S. Department of Energy, Office of Science, Nuclear Physics program office under Award Number DE-SC0015136.

\paragraph{Disclaimer}  This publication was prepared as an account of work sponsored by an agency of the United States Government. Neither the United States Government nor any agency thereof, nor any of their employees, makes any warranty, express or implied, or assumes any legal liability or responsibility for the accuracy, completeness, or usefulness of any information, apparatus, product, or process disclosed, or represents that its use would not infringe privately owned rights. Reference herein to any specific commercial product, process, or service by trade name, trademark, manufacturer, or otherwise does not necessarily constitute or imply its endorsement, recommendation, or favoring by the United States Government or any agency thereof. The views and opinions of authors expressed herein do not necessarily state or reflect those of the United States Government or any agency thereof.

\bibliographystyle{ieeetr}
\bibliography{refs}

\begin{thebibliography}{10}

\bibitem{det_matrixx_1}
M.~V. Anvar, A.~Attili, M.~Ciocca, M.~Donetti, L.~F. Guarachi, F.~Fausti,
  S.~Giordanengo, F.~Marchetto, S.~Molinelli, V.~Monaco, {\em et~al.},
  ``{Quality assurance of carbon ion and proton beams: a feasibility study for
  using the 2D MatriXX detector},'' {\em Physica Medica}, vol.~32, no.~6,
  pp.~831--837, 2016.

\bibitem{stelljes2015dosimetric}
T.~Stelljes, A.~Harmeyer, J.~Reuter, H.~Looe, N.~Chofor, D.~Harder, and
  B.~Poppe, ``{Dosimetric characteristics of the novel 2D ionization chamber
  array OCTAVIUS Detector 1500},'' {\em Medical physics}, vol.~42, no.~4,
  pp.~1528--1537, 2015.

\bibitem{shen2017using}
J.~Shen, E.~Tryggestad, J.~E. Younkin, S.~R. Keole, K.~M. Furutani, Y.~Kang,
  M.~G. Herman, and M.~Bues, ``Using experimentally determined proton spot
  scanning timing parameters to accurately model beam delivery time,'' {\em
  Medical physics}, vol.~44, no.~10, pp.~5081--5088, 2017.

\bibitem{kohno2017development}
R.~Kohno, K.~Hotta, T.~Dohmae, Y.~Matsuzaki, T.~Nishio, T.~Akimoto,
  T.~Tachikawa, T.~Asaba, J.~Inoue, T.~Ochi, {\em et~al.}, ``Development of
  continuous line scanning system prototype for proton beam therapy,'' {\em
  International Journal of Particle Therapy}, vol.~3, no.~4, pp.~429--438,
  2017.

\bibitem{pidikiti2018commissioning}
R.~Pidikiti, B.~C. Patel, M.~R. Maynard, J.~P. Dugas, J.~Syh, N.~Sahoo, H.~T.
  Wu, and L.~R. Rosen, ``Commissioning of the world's first compact pencil-beam
  scanning proton therapy system,'' {\em Journal of applied clinical medical
  physics}, vol.~19, no.~1, pp.~94--105, 2018.

\bibitem{fraser1979beam}
J.~Fraser, ``Beam analysis tomography,'' {\em IEEE Transactions on Nuclear
  Science}, vol.~26, no.~1, pp.~1641--1645, 1979.

\bibitem{svensson2011non}
J.~Svensson, {\em Non-parametric tomography using Gaussian processes}.
\newblock EFDA, 2011.
\newblock
  \url{http://www.euro-fusionscipub.org/wp-content/uploads/eurofusion/EFDP11024.pdf}.

\bibitem{li2013bayesian}
D.~Li, J.~Svensson, H.~Thomsen, F.~Medina, A.~Werner, and R.~Wolf, ``Bayesian
  soft {X}-ray tomography using non-stationary {Gaussian} {Processes},'' {\em
  Review of Scientific Instruments}, vol.~84, no.~8, p.~083506, 2013.

\bibitem{wang2018gaussian}
T.~Wang, D.~Mazon, J.~Svensson, D.~Li, A.~Jardin, and G.~Verdoolaege,
  ``Gaussian process tomography for soft {X}-ray spectroscopy at {WEST} without
  equilibrium information,'' {\em Review of Scientific Instruments}, vol.~89,
  no.~6, p.~063505, 2018.

\bibitem{rdi_patent_one}
E.~Galyaev, ``Ionizing particle beam fluence and position detector array using
  micromegas technology with multi-coordinate readout,'' Apr.~23 2019.
\newblock US Patent App. 10/265,545.

\bibitem{lawson1995solving}
C.~L. Lawson and R.~J. Hanson, {\em Solving least squares problems}, vol.~15.
\newblock Siam, 1995.

\bibitem{chen2010nonnegativity}
D.~Chen and R.~J. Plemmons, ``Nonnegativity constraints in numerical
  analysis,'' in {\em The birth of numerical analysis}, pp.~109--139, World
  Scientific, 2010.

\bibitem{Palladino1975drift}
V.~Palladino and B.~Sadoulet, ``Application of classical theory of electrons in
  gases to drift proportional chambers,'' {\em Nuclear Instruments and
  Methods}, vol.~128, no.~2, pp.~323--335, 1975.

\bibitem{finger1985hexagonal}
M.~Finger and T.~Prince, ``Hexagonal uniformly redundant arrays for
  coded-aperture imaging,'' in {\em International Cosmic Ray Conference},
  vol.~3, 1985.

\bibitem{nngptRepo}
``Nonnegative gaussian process tomography code repository.''
  \url{https://github.com/decibelcooper/nngpt}.
\newblock Accessed: 2019-12-1.

\end{thebibliography}

\end{document}